\documentclass[aps,pre,amsmath,amssymb,reprint,showpacs] {revtex4-1}

\usepackage{graphicx}
\usepackage{dcolumn}
\usepackage{bm}
\usepackage{ifthen}
\usepackage{psfrag}
\usepackage{booktabs}

\newcommand{\diff}[2]{\frac{\partial #1}{\partial #2}}

\newcommand{\avar}[1]{\langle #1 \rangle}
\newcommand{\klamm}[1]{\left( #1 \right)}
\newcommand{\norm}[1]{\left\| #1 \right\|}

\newcommand{\istobe}{{\overset{!}=}}

\newcommand{\ofR}{\klamm{\bf r}}
\newcommand{\ofRD}{\klamm{\bf r'}}
\newcommand{\newstuff}[1]{{{ #1}}}
\newcommand{\defi}{{\mathrel{\mathop:}=}}

\newcommand{\SIA}{{\text{SIA}}}
\newcommand{\SKA}{{\text{SKA}}}
\newcommand{\Sph}{{\text{Sph}}}
\newcommand{\Pla}{{\text{Pla}}}
\newcommand{\HS}{{\text{HS}}}

\newcommand{\Offiziell}{1}
\newcommand{\Comment}[1]{ \ifthenelse{\Offiziell = 0}{{\bf #1}}{}}

\newcommand{\bb}{\begin{equation}}
\newcommand{\ee}{\end{equation}}
\newcommand{\ba}{\begin{eqnarray}}
\newcommand{\ea}{\end{eqnarray}}
\newcommand{\rhor}{\rho({\bf r})}

\newcommand{\rr}{{\mathbf r}}
\newcommand{\dr}{{\rm d}{\bf r}}

\usepackage{natbib}

\begin{document}

\preprint{AIP/123-QED}

\title{Wetting on a spherical wall: influence of liquid-gas interfacial properties}

\author{Andreas Nold}
\email{andreas.nold09@imperial.ac.uk}
\affiliation{Center of Smart
Interfaces, TU Darmstadt, Petersenstr. 32, 64287 Darmstadt, Germany \\
Department of Chemical Engineering, Imperial College London, London SW7 2AZ, United Kingdom}

\author{Alexandr Malijevsk\'y}
\email{a.malijevsky@imperial.ac.uk}
\affiliation{E. H\'ala
Laboratory of Thermodynamics, Institute of Chemical Process
Fundamentals of ASCR, 16502 Prague 6, Czech Republic}
\affiliation{Department of Physical Chemistry, Institute of Chemical Technology, Prague, 166 28 Praha 6, Czech Republic}%

\author{Serafim Kalliadasis}
\email{s.kalliadasis@imperial.ac.uk}
\affiliation{Department of Chemical Engineering, Imperial College London, London SW7 2AZ, United Kingdom}

\date{\today}

\begin{abstract}
We study the equilibrium of a liquid film on an attractive spherical substrate for an intermolecular interaction
model exhibiting both fluid-fluid and fluid-wall long-range forces. We first reexamine the wetting properties of
the model in the zero-curvature limit, i.e., for a planar wall, using an effective interfacial Hamiltonian approach
in the framework of the well known sharp-kink approximation (SKA). We obtain very good agreement with
a mean-field density functional theory (DFT), fully justifying the use of SKA in this limit. We then turn our
attention to substrates of finite curvature and appropriately modify the so-called soft-interface approximation
(SIA) originally formulated by Napi\'orkowski and Dietrich [Phys. Rev. B 34, 6469 (1986)] for critical wetting
on a planar wall. A detailed asymptotic analysis of SIA confirms the SKA functional form for the film growth.
However, it turns out that the agreement between SKA and our DFT is only qualitative. We then show that the
quantitative discrepancy between the two is due to the overestimation of the liquid-gas surface tension within
SKA. On the other hand, by relaxing the assumption of a sharp interface, with, e.g., a simple ÒsmoothingÓ of
the density profile there, markedly improves the predictive capability of the theory, making it quantitative and
showing that the liquid-gas surface tension plays a crucial role when describing wetting on a curved substrate.
In addition, we show that in contrast to SKA, SIA predicts the expected mean-field critical exponent of the
liquid-gas surface tension.
\end{abstract}


\maketitle

\section{Introduction} \label{Intro}

The behavior of fluids in confined geometries, in particular in the
vicinity of solid substrates, and associated wetting phenomena are
of paramount significance in numerous technological applications and
natural phenomena. Wetting is also central in several fields, from
engineering and materials science to chemistry and biology. As a
consequence, it has received considerable attention, both
experimentally and theoretically for several decades. Detailed and
comprehensive reviews are given in
Refs.~\cite{Dietrich,RevModPhys.81.739,Schick1990LesHouches,SullivanDaGama1986}.

Once a substrate (e.g. a solid wall) is brought into contact with a gas, the substrate-fluid attractive forces cause adsorption of some of the fluid molecules on the substrate surface, such that at least a microscopically thin
liquid film forms on the surface. The interplay between the fluid-fluid interaction (cohesion) and the fluid-wall interaction (adhesion) then determines a particular wetting state of the system. This state can be quantified by
the contact angle at which the liquid-gas interface meets the substrate. If the contact angle is \newstuff{non-zero}, i.e. a spherical cap of the liquid is formed on the substrate, the surface is called partially wet. In the regime of
partial wetting, the cap is surrounded by a thin layer of adsorbed fluid which is of molecular dimension. Upon approaching the critical temperature, the contact angle continuously decreases and eventually vanishes. Beyond this
wetting temperature one speaks of complete wetting and the film thickness becomes of macroscopic dimension. The transition between the two regimes can be qualitatively distinguished by the rate of the disappearance of the
contact angle, which is discontinuous in the case of a first-order transition or continuous for critical wetting.

From a theoretical point of view, it is much more convenient to take
the adsorbed film thickness, $\ell$, rather than the contact angle,
as an order parameter for wetting transitions and related phenomena. An
interfacial Hamiltonian is then minimized with respect to $\ell$ as
is typically the case with the (mesoscopic) Landau-type field
theories and (microscopic) density functional theory (DFT) -- where
$\ell$ can be easily determined from the Gibbs adsorption, a direct
output of DFT.

In this study, we examine the wetting properties of a simple fluid
in contact with a spherical attractive wall by using an
intermolecular interaction model with fluid-fluid and fluid-wall
long-range forces. The curved geometry of the system prohibits a
macroscopic growth of the adsorbed layer (and thus complete
wetting), since the free energy contribution due to the liquid-gas
interface increases with the film thickness $\ell$, and thus for a
given radius of a spherical substrate there must be a maximum finite
value of
$\ell$~\cite{Dietrich,DietrichWettingOnCurvedSubstrates,Holstrokyst:1987fk}.
For the mesoscopic approaches, the radius of the wall $R$, is a new
field variable that introduces one additional $\ell$-dependent term
to the effective interface Hamiltonian of the system, compared to
the planar geometry, where the only $\ell$-dependent term is the
binding potential between the wall-liquid and liquid-gas interfaces.
Furthermore, for a fluid model exhibiting a gas-liquid phase
transition, such as ours, it has been found that two regimes of the
interfacial behavior should be distinguished:  $R>R_C$, in which
case the surface tension can be expanded in integer powers of
$R^{-1}$ and $R<R_C$, where the interfacial quantities exhibit a
non-analytic behavior~\cite{Evans:2004ys}. Moreover, for an
intermolecular interaction model with fluid-fluid long-range
interactions, there is an additional $R^{-2}\log R$ contribution to
the surface tension in the $R>R_C$ regime~\cite{StewartEvans}. These
striking observations actually challenge all curvature expansion
approaches. In addition, a certain equivalence between a system of a
saturated fluid on a spherical wall and a system of an unsaturated
fluid on a planar wall above the wetting temperature has been
found~\cite{DietrichWettingOnCurvedSubstrates,StewartEvans}.
Somewhat surprisingly, DFT computations confirmed this
correspondence at the level of the density profiles down to
unexpectedly small radii of the wall~\cite{StewartEvans}.

Most of these conjectures follow from the so-called sharp-kink
approximation (SKA)~\cite{Dietrich}, based on a simple piece-wise
constant approximation of a one-body density distribution of the
fluid, i.e. a coarse-grained approach providing a link between
mesoscopic Hamiltonian theories and microscopic DFT. The simple
mathematical form of SKA has motivated many theoretical
investigations of wetting phenomena as it makes them analytically
tractable. At the same time SKA appears to capture much of the
underlying fundamental physics for planar substrates (often in
conjugation with exact statistical mechanical sum
rules~\cite{Henderson1992}).

\newstuff{However, as we show in this work, SKA is only qualitative
for spherical substrates, even though the functional form of the
film growth can still be successfully inferred from the
theory~\cite{StewartEvans}.
We attribute this to the particular approximation of the liquid-gas
interface adapted by SKA. In particular, since the $\ell$-dependent
contribution to the interface Hamiltonian due to the curvature is
proportional to the liquid-gas surface tension, the latter plays an
important role compared to the planar geometry.}

More
specifically, the curved geometry induces a Laplace pressure whose
value depends on both film thickness and the surface tension and so
the two quantities are now coupled, in contrast with the planar
geometry where a parallel shift of the liquid-gas dividing surface
does not influence the surface contribution to the free energy of
the system.
We further employ an alternative coarse-grained approach, a
modification of the one originally proposed by Napi\'orkowski and
Dietrich~\cite{DietrichNapiorkowski_BulkCorrelation} for the planar
geometry, which replaces the jump in the density profile at the
liquid-gas interface of SKA by a continuous function restricted by
several reasonable constraints. We show that in this
``soft-interface approximation" (SIA) the leading curvature
correction to the liquid-gas surface tension is $O(R^{-1})$, rather
than $O(R^{-2}\log R)$, in line with the Tolman theory. Once a
particular approximation for the liquid-gas interface is taken, the
corresponding Tolman length can be easily determined. Apart from
this, we find that the finite width of the liquid-gas interface
significantly improves the prediction of the corresponding surface
tension when compared with the microscopic DFT computations, which
consequently markedly improves the estimation of the film thickness
in a spherical geometry.

In Sec.~\ref{sec:DFT} we describe our microscopic model and the
corresponding DFT formalism. In
Sec.~\ref{sec:WettingOnAPlanarSubstrate} we present results of
wetting phenomena on a planar wall obtained from our DFT based on a
continuation scheme that allows us to trace metastable and unstable
solutions. The results are compared with the analytical prediction
as given by a minimization of the interface Hamiltonian based on
SKA. We also make a connection between the two approaches by
introducing the microscopic model into the interfacial Hamiltonian.
In Sec. \ref{sec:WettingCurvedSubstrate} we turn our attention to
the main part of our study, a thin liquid film on a spherical wall.
\newstuff{We show that the SKA does not perform as well as might be desired, in
particular, it does not account for a quantitative description of
the liquid-gas surface tension which plays a significant role when
the substrate geometry is curved.} We then introduce SIA and present
an asymptotic analysis with the new approach. Comparison with DFT
computations reveals a substantial improvement of the resulting
interface Hamiltonian even for very simple approximations of the
density distribution at the liquid-vapour interface, indicating the
significance of a non-zero width of the interface. We conclude in
Sec.~\ref{sec:summary} with a summary of our results and discussion.
Appendix A describes the continuation method we developed for the
numerical solution of DFT. In Appendix B we show derivations of the
surface tension and the binding potential for both a planar and a
spherical geometry within SKA. Finally, Appendix C shows derivations
of the above quantities, including Tolman's length, using SIA.

\section{DFT \label{sec:DFT}}
\subsection{General formalism}
DFT is based on Mermin's proof \cite{Mermin:1965fk} that the free
energy of an inhomogeneous system at equilibrium can be expressed as
a functional of an ensemble averaged one-body density, $\rhor$ (see
e.g. Ref.~\cite{Evans} for more details). Thus, the free-energy
functional $\mathcal{F}[\rho]$ contains all the equilibrium physics
of the system under consideration. Clearly, for a 3D fluid model one
has to resort to an approximative functional. Here we adopt a simple
but rather well established local density approximation
 \begin{align}
 \mathcal{F}[\rho] =& \int f_{\HS}\klamm{\rho\ofR} \rho \ofR d{\bf r} + \notag\\
&+\frac{1}{2}\iint \rhor\rho(\rr')\phi\klamm{|{\bf r} - {\bf r}'|} d{\bf r}' d{\bf r}, \label{eq:FreeEnergySplitting}
\end{align}
where $f_{\HS}\klamm{\rho\ofR}$ is the free energy per particle of
the hard-sphere fluid (accurately described by the Carnahan-Starling
equation of state), including the ideal gas contribution. The
contribution due to the long-range van der Waals forces is included
in the mean-field manner. To be specific, we consider a full
Lennard-Jones 12-6 (LJ) potential to model the fluid-fluid
attraction according to the Barker-Henderson perturbative scheme
\begin{equation}
\phi\klamm{r} =  \left\{\begin{array}{ll}
0 & r < \sigma\\
4 \varepsilon\klamm{\klamm{\frac{\sigma}{r}}^{12} -
\klamm{\frac{\sigma}{r}}^6} & r \geq \sigma
\end{array}\right.\,,
\label{eq:Definition_PhiP}
\end{equation}
where for the sake of simplicity the Lennard-Jones parameter $\sigma$ is taken equal to the hard-sphere diameter.

The free-energy functional, $\mathcal{F}[\rho]$, describes the
intrinsic properties of a given fluid. The total free energy
including also a contribution of the external field is related to
the grand potential functional through the Legendre transform
\begin{align}
\Omega[\rho] = \mathcal{F}[\rho] + \int \rho\ofR \klamm{V\ofR - \mu } d{\bf r}, \label{eq:GrandPotentialOne}
\end{align}
where $\mu$ is the chemical potential and $V\ofR$ is the external field due to the presence of a wall $W \subset \mathbb{R}^3$,
\begin{align}
V\ofR =  \left\{\begin{array}{ll}
\infty &  {\bf r} \in W\\
\rho_{w} \int_W \phi_w\klamm{|{\bf r}-{\bf r}'|} d{\bf r}' &
\text{ elsewhere, }
\end{array}\right.
\label{eq:Definition_WallPotential}
\end{align}
consisting of the atoms interacting with the fluid particles via the
Lennard-Jones potential, $\phi_w\klamm{r}$, with the parameters
$\sigma_w$ and $\varepsilon_w$, and uniformly distributed throughout
the wall with a density $\rho_w$:
\begin{align}
\phi_{w}\klamm{r} =
4 \varepsilon_{w}\klamm{\klamm{\frac{\sigma_{w}}{r}}^{12} -
\klamm{\frac{\sigma_{w}}{r}}^6} .
\end{align}

Applying the variational principle to the grand potential
functional, Eq.~(\ref{eq:GrandPotentialOne}), we attain the
Euler-Lagrange equation:
 \begin{align}
\frac{\delta \mathcal{F}_{\HS}[\rho]}{\delta \rho\ofR} + \int \rho\ofRD \phi\klamm{|{\bf r} - {\bf r}'|} d{\bf r}'
 + V\ofR - \mu =0, \label{eq:ExtremalCondition}
\end{align}
where $\mathcal{F}_{\HS}[\rho]$ denotes the first term in the
right-hand-side of (\ref{eq:FreeEnergySplitting}). In general, the
solution of (\ref{eq:ExtremalCondition}) comprises all extremes of
the grand potential
 $\Omega[\rho]$ as given by (\ref{eq:GrandPotentialOne}) and
not just the global minimum corresponding to the equilibrium state.
Here we develop a
pseudo arc-length continuation scheme for
the numerical computation of (\ref{eq:ExtremalCondition}) that
enables us to capture both locally stable and unstable solutions and
thus to construct the entire bifurcation diagrams for the isotherms
(details of the scheme are given in Appendix
\ref{sec:Annay:NumericalMethod}).

\begin{figure}[hbt]
\includegraphics[width=8.5cm]{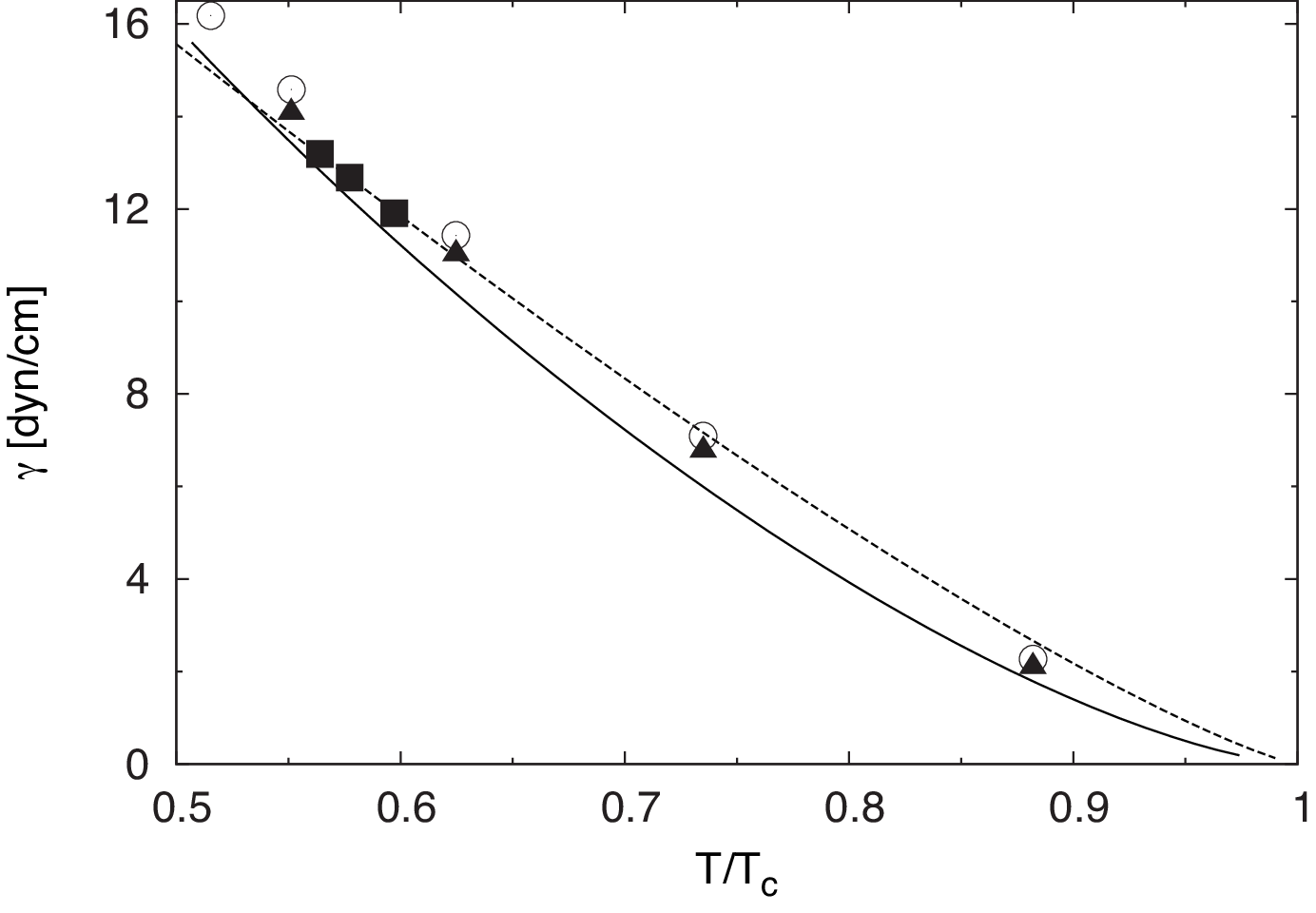}
\caption{Plots of surface tension as a function of dimensionless temperature,
$T/T_c$. Solid line: numerical DFT results of our model scaled with $\varepsilon/k_{B}=119.8K$ and $\sigma = 3.4 \mathring{A}$;
triangles: computational results by Toxvaerd for a 12-6 LJ fluid
using the Barker-Henderson perturbation theory~\cite{BarkerHendersonD} with the Percus-Yevick solution~\cite{Throop} for the hard-sphere reference fluid and using the exact
hard sphere diameter~\cite{Toxvaerd}; circles: Monte Carlo
simulations by Lee and Barker~\cite{LeeBarker}; squares:
experimental results for Argon by Guggenheim~\cite{Guggenheim};
dashed line: fit of experimental results to
equation $\gamma(T) = \gamma_0 \klamm{1- {T}/{T_C}}^{1 + r}$ by
Guggenheim~\cite{Guggenheim}. The resulting coefficients are
$\gamma_0 = 36.31 \text{dyn}/\text{cm}$ and $r = \frac{2}{9}$.
\label{fig:SurfaceTension}}
\end{figure}

The excess part of the grand potential functional
(\ref{eq:GrandPotentialOne}) over the bulk may be expressed in the
form
\begin{align}
\Omega_{ex}[\rho\ofR] =&
-\int \klamm{p\klamm{\rho\ofR} - p\klamm{\rho_{b}}} d{\bf r}+\notag\\
+& \frac{1}{2} \iint \rho\ofR \klamm{ \rho\ofRD - \rho\ofR }\phi(|\bf r'-\bf r|) d{\bf r'} d{\bf r}+\notag\\
+&\int \rho\ofR V\ofR d{\bf r},
\label{eq:DefinitionExcessGrandPotential}
\end{align}
where $\rho_{b}$ is the density of the bulk phase and
\begin{align}
- p\klamm{\rho}= \rho f_{\HS}\klamm{\rho} + \alpha \rho^2 - \mu \rho, \label{eq:DefinitionPressure}
\end{align}
is the negative pressure, or grand potential per unit volume, of a
system with uniform density $\rho$ and $\alpha \equiv
\frac{1}{2}\int \phi\klamm{|{\bf r}|} d{\bf r} = -\frac{16}{9}\pi
\varepsilon \sigma^{3}$. In particular, the equilibrium value of the
excess grand potential (\ref{eq:DefinitionExcessGrandPotential}) per
unit area of a two-phase system of liquid and vapour in the absence
of an external field, yields the surface tension between the
coexisting phases, $\gamma_{lg}$. The prediction of $\gamma_{lg}$ as
given by minimization of (\ref{eq:DefinitionExcessGrandPotential})
agrees fairly well with both computations and experimental data as
shown in Fig.~\ref{fig:SurfaceTension}.

\subsection{Translational symmetry: planar wall} \label{subsec:planar}
If the general formalism outlined above is applied on a particular
external field attaining a certain symmetry, it will adopt a
significantly simpler form. In the next subsection we will formulate
the basic equations resulting from the equilibrium conditions
obtained from the minimization
of~(\ref{eq:DefinitionExcessGrandPotential}), for a spherical model
of the external field, i.e. a system with rotational symmetry. But
prior to that, it is instructive to discuss the zero-curvature limit
of the above model, corresponding to an adsorbed LJ fluid on a
planar wall, a system with translational symmetry.

For a planar substrate $W = \mathbb{R}^2\times \mathbb{R}^-$ in
Cartesian coordinates, the density profile is only a function of
$z$, so that the Euler-Lagrange equation reads
\begin{align}
\mu_{\HS}\klamm{\rho(z)} + \int_{0}^\infty \rho(z') \Phi_{\Pla}\klamm{|z - z'|} dz' +&\label{eq:ExtremalCondition1D}\\
 + V_{\infty}(z) - \mu = 0 &\quad \klamm{\forall z \in \mathbb{R}^+},\notag
\end{align}
where $\mu_{\HS}(\rho)= \diff{\klamm{f_{\HS}(\rho)\rho}}{\rho}$ is
the chemical potential of the hard-sphere system.
A fluid particle at a distance $z$ from the wall experiences the
wall potential:
\begin{align}
V_{\infty}(z) &=
\rho_{w} \int_{W} \phi_{w}(\sqrt{x'^{2} + y'^{2} + (z-z')^{2}}) dx' dy' dz'  \notag\\
&=
\left\{\begin{array}{ll}
\infty &   z \leq 0\\
4\pi \rho_w\varepsilon_w  \sigma_w^3\klamm{ \frac{1}{45}\klamm{\frac{\sigma_w}{ z}}^9 - \frac{1}{6}\klamm{\frac{\sigma_w}{ z}}^3 } & z > 0.
\end{array}\right.  \label{eq:ExternalPot_Planar}
\end{align}
$\Phi_{\Pla}(z)$ in Eq.~(\ref{eq:ExtremalCondition1D}) is the
surface potential exerted by the fluid particles uniformly
distributed (with a unit density) over the $x$-$y$ plane at distance
$z$:
\begin{align}
\Phi_{\Pla}(z) &= \iint \phi\klamm{ \sqrt{x^2 + y^2 + z^2} } dy dx \notag\\
&= 2 \pi \int_{0}^{\infty} \phi\klamm{ \sqrt{z^{2} + r^{2}  } } r dr \label{eq:Definition_Phi1D}\\
&= -\frac{6}{5} \pi\varepsilon\sigma^2\times
 \left\{\begin{array}{ll}
1 & z < \sigma\,,\\
\frac{5}{3}
\klamm{\frac{\sigma}{z}}^{4}
-
\frac{2}{3}
\klamm{ \frac{\sigma}{ z} }^{10}
 & z \geq \sigma\,.
\end{array}\right. \notag
\end{align}

In the framework of DFT, the natural order parameter for wetting
transitions is the Gibbs adsorption per unit area:
\begin{align}
\Gamma_\infty[\rho\klamm{z}] = \int_0^\infty \klamm{\rho(z) -
\rho_{b}} dz.
\end{align}
\subsection{Rotational symmetry: spherical wall} \label{sphere}
If the external field is induced by a spherical wall, $W = \{{\bf r}
\in \mathbb{R}^3: r\equiv|{\bf r}|< R \}$, the variational principle
yields
\begin{align}
\mu_{\HS}\klamm{\rho(r)} + \int_R^\infty \rho(r') \Phi_{\Sph}\klamm{r,r'} dr' +&\label{eq:ExtremalConditionSph}\\
 + V_{R}(r) - \mu &= 0, \quad \klamm{\forall r > R},\notag
\end{align}
where $\Phi_{\Sph}\klamm{r,r'}$ is the surface interaction potential
per unit density generated by fluid particles uniformly distributed
on the surface of the sphere $B_{r'}$ centered at the origin at
distance $r$,
\begin{align}
\Phi_{\Sph}\klamm{r,r'} &= \int_{\partial B_{r'}} \phi\klamm{|{\bf
r} - {\bf \tilde r}|}d{\bf \tilde r}. \label{eq:Definition_PhiSph}\\
&= \frac{ r'}{ r} \klamm{ \Phi_{\Pla}\klamm{| r- r'|} -
\Phi_{\Pla}\klamm{| r +  r'|}}\notag
\end{align}
(see also Appendix \ref{sex:AppendisSKASurfaceTension}). The wall
potential in Eq.~(\ref{eq:Definition_WallPotential}) for the
spherical wall $W = \{{\bf  r} \in \mathbb{R}^3: |{\bf  r}| \leq
R\}$ is:
\begin{align}
 V_{R}( r) = &\frac{\rho_w\varepsilon_w
\sigma_w^4 \pi }{3 r}\left\{
\frac{\sigma_w^8}{30}\left[ \frac{ r+9 R}{(
r+ R)^9} - \frac{ r-9 R}{( r- R)^9}\right]
+ \right.\notag\\
&+\left. \sigma_w^2\left[ \frac{ r-3 R}{(
r- R)^3} - \frac{ r+3 R}{( r+ R)^3}
\right]\right\}. \label{eq:Annex_ExternalPot_Spherical}
\end{align}
Replacing the distance from the origin $r$ by the radial distance from the wall $\tilde r = r - R$, one can easily see
that the external potential (\ref{eq:Annex_ExternalPot_Spherical})
reduces to the planar wall potential (\ref{eq:ExternalPot_Planar}), for
$R\to \infty$.
Analogously to the planar case, we define the
adsorption $\Gamma_R$ as the excess number of particles of the
system with respect to the surface of the wall:
\begin{align}
\Gamma_{R}[\rho(r)] = \int_R^\infty \klamm{\frac{r}{R}}^2 \klamm{\rho(r) -
\rho_{b}} dr.
\end{align}
\section{Wetting on a planar substrate \label{sec:WettingOnAPlanarSubstrate}}
In this section we make a comparison between the numerical solution
of DFT and the prediction given by the effective interfacial
Hamiltonian according to SKA for the first-order wetting transition
on the planar substrate. We consider a planar semi-infinite wall
interacting with the fluid according to
(\ref{eq:ExternalPot_Planar}) with the \newstuff{typical parameters
$\rho_w\varepsilon_w=0.8 \varepsilon/\sigma^{3}$ and $\sigma_w=1.25
\sigma$ that correspond to the class of intermediate-substrate
systems \cite{PhysRevB.26.5112} for which prewetting phase
transitions can be observed. We note that wetting on planar and
spherical walls is a multiparametric problem and hence a full
parametric study of the global phase diagram is a difficult task,
beyond the scope of this paper. }
\subsection{Numerical DFT results of wetting on a planar wall \label{sec:PlanarIsotherm}}
\begin{figure}
\includegraphics[width=8.5cm]{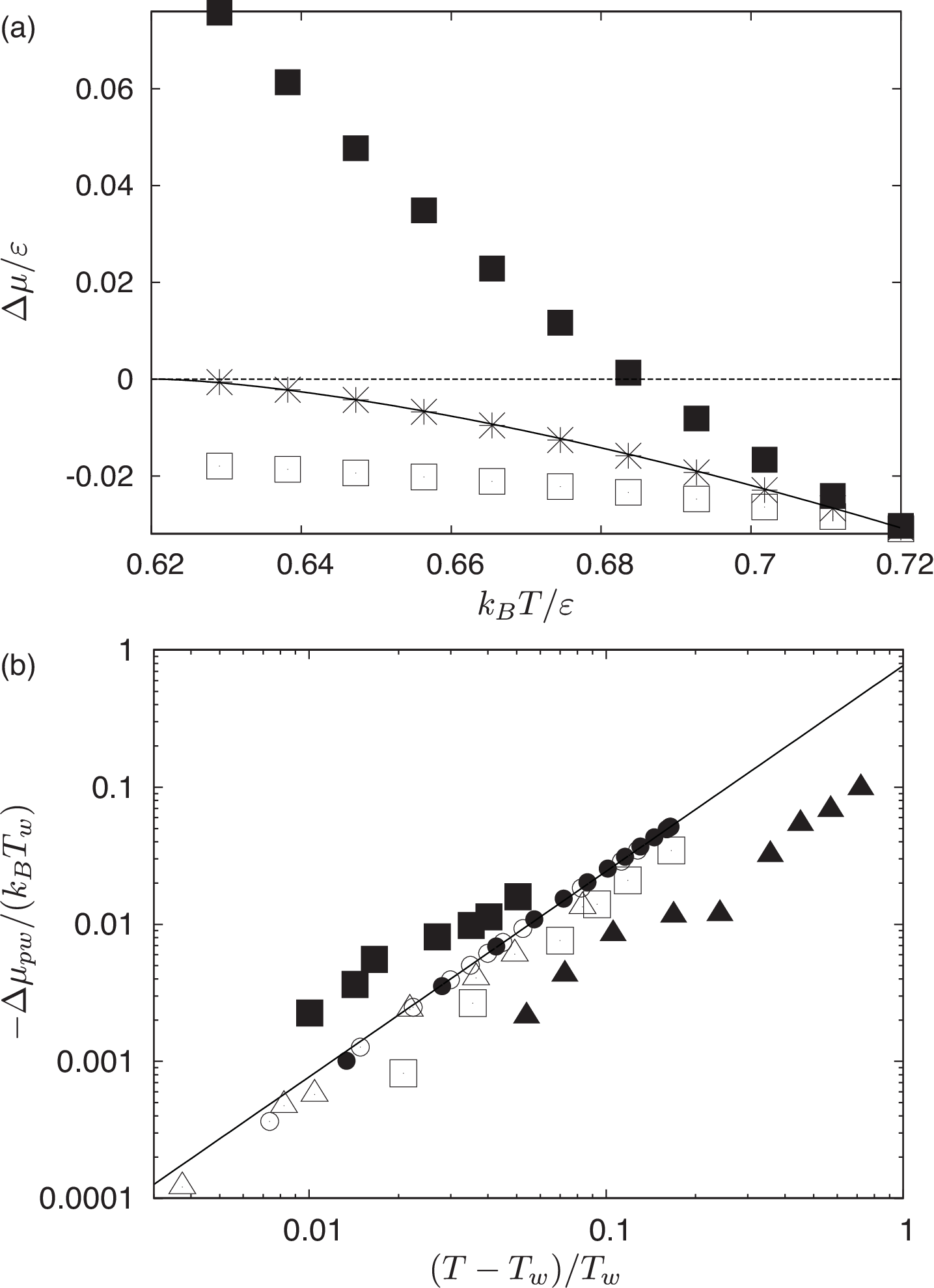}
\caption{
(a) The deviation of the chemical potential from its
saturation value at prewetting (crosses), and at the left (open squares)
and right (filled squares) saddle nodes of bifurcation as a
function of temperature. The dashed line marks the locus of the
chemical potential at saturation for the given temperature,
$\Delta \mu = 0$.
The solid line is a fit to $- \Delta \mu_{pw}(T)/(k_{B}T_{w}) = C((T -T_w)/T_{w})^{3/2}$
where the wetting temperature is $k_{B}T_{w} = 0.621\varepsilon$ and the prewetting critical temperature is
$k_{B}T_{pwc} =0.724 \varepsilon$. The resulting
coefficient is $C=0.77$.
(b) The scaled prewetting phase diagrams for different systems.
The circles are DFT calculations for an attractive wall with
$\sigma_{w}= 1.25\sigma$ and  $\rho_w\varepsilon_w=0.8 \varepsilon/\sigma^3$ (open circles) and
$\rho_w\varepsilon_w=0.75 \varepsilon/\sigma^3$ (filled circles).
Experimental data~\cite{BonnRoss}: filled squares, methanol on cyclohexane (Kellay
et. al. 1993)~\cite{Kellay}; open triangles, ${\rm H}_{2}$ on rubidium
(Mistura \emph{et al.} 1994) \cite{Mistura};
filled triangles, He on caesium (Rutledge \emph{et al.}, 1997)~\cite{RutledgeTaborek1992};
open squares, ${\rm H}_{2}$ on Caesium (Ross \emph{et al.}, 1997)~\cite{Ross}.
\label{fig:MuNodd_Temp}
}
\end{figure}

Figure~\ref{fig:MuNodd_Temp} depicts the surface-phase diagram  of
the considered model in the $(\Delta\mu,T)$ plane, where $\Delta\mu
= \mu - \mu_{sat}$ is the departure of the chemical potential from
its saturation value. The first-order wetting transition takes place
at wetting temperature $k_BT_w=0.621 \varepsilon $, well bellow the
critical temperature of the bulk fluid $k_{B}T_c=1.006\varepsilon$
for our model. The prewetting line connects the saturation line at
the wetting temperature $T_w$ and terminates at the prewetting
critical point, $k_BT_{pwc}=0.724 \varepsilon$. The slope of the
prewetting line is governed by a Clapeyron-type
equation~\cite{HaugeSchick}, which, in particular, states that the
prewetting line approaches the saturation line tangentially at $T_w$
with
\begin{align}
\left. \frac{d\klamm{\Delta \mu_{pw}}}{dT}\right|_{T=T_w} = 0,
\end{align}
in line with our numerical computations. Schick and
Taborek~\cite{SchickTaborek} later showed that the prewetting line
scales as $-\Delta \mu \sim \klamm{T-T_w}^{3/2}$. In
Ref.~\cite{BonnRoss}, this power law was confirmed experimentally,
such that
\begin{align}
- \frac{ \Delta \mu_{pw}(T)}{k_{B}T_{w}} = C \klamm{ \frac{T-T_{w}}{T_{w}}}^{3/2}, \label{eq:PrewettingLaw}
\end{align}
with $C\approx \frac{1}{2}$~\cite{BonnRoss}. A fit of our DFT
results with (\ref{eq:PrewettingLaw}) leads to a coefficient
$C=0.77$, in a reasonable agreement with the experimental data --
see Fig. \ref{fig:MuNodd_Temp}.

\begin{figure}[hbt]
\includegraphics[width=8.0cm]{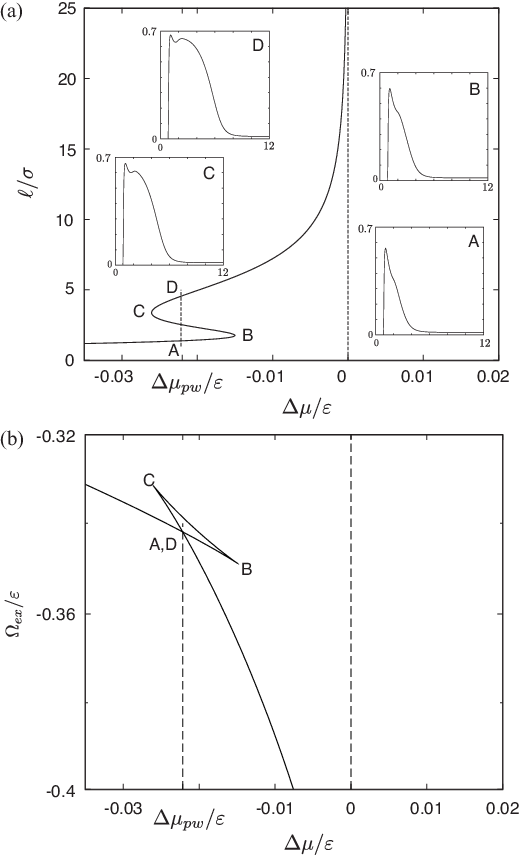}
\caption{\label{fig:IsothermProfiles}
(a) The $\ell$--$\Delta \mu$
bifurcation diagram for $k_{B}T= 0.7\varepsilon$ for a wall with
$\rho_w\varepsilon_w  = 0.8 \varepsilon/\sigma^{3}$ and $\sigma_w = 1.25 \sigma$.
$\Delta \mu$ is the deviation of the chemical potential from its
saturation value, $\mu_{sat}$. The prewetting transition, marked by
the dashed line, occurs at chemical potential $\Delta \mu_{pw} = -0.022\varepsilon$.
The inset subplots show the density $\rho\sigma^{3}$ as a function of
the distance $z/\sigma$ from the wall. (b) The excess grand potential
$\Omega_{ex}/\varepsilon$ as a function of $\Delta \mu /
\varepsilon$ in the vicinity of the prewetting transition.}
\label{fig:PlanarIsotherm}
\end{figure}

Figure~\ref{fig:IsothermProfiles} depicts the adsorption isotherm in
terms of the thickness of the adsorbed liquid film $\ell$ as a
function of $\Delta \mu$ for the temperature $k_BT=0.7 \varepsilon$
and in the interval between the wetting temperature $T_w$ and the
prewetting critical temperature $T_{pwc}$. $\ell$ can be associated
with the Gibbs adsorption through
\begin{align}
\ell = \frac{\Gamma_R[\rho]}{\Delta \rho},
\end{align}
for both finite and infinite $R$, where
$\Delta\rho=\rho_{l}^{sat}-\rho_{g}^{sat}$ is the difference between
the liquid and gas densities at saturation.

The isotherm exhibits a van der Waals loop with two turning points
depicted as B and C demarcating the unstable branch. Points A and D
indicate the equilibrium between thin and thick layers,
corresponding to a point on the prewetting line in
Fig.~\ref{fig:MuNodd_Temp}. The location of the equilibrium points
can be obtained from a Maxwell construction. Details of the
numerical scheme we developed for tracing the adsorption isotherms
are given in Appendix~\ref{sec:Annay:NumericalMethod}.

\subsection{SKA for a planar wall  \label{sec:PlanarSharpInterface}}
For the sake of clarity and completeness we briefly review the main
features of SKA for a planar geometry (details are given in
Ref.~\cite{Dietrich}).

Let us consider a liquid film of a thickness $\ell$ adsorbed on a
planar wall. According to the SKA the density distribution is
approximated by a piecewise constant function
\begin{align}
\rho^{\SKA}_{\ell}(z) =  \left\{\begin{array}{ll}
0 &  z<\delta\,,\\
\rho_{l}^{+} & \delta<z<\ell\,,\\
\rho_{g} &  z>\ell\,,
\end{array}\right. \label{eq:Sharp-Interface-Profile1D}
\end{align}
where $\rho_{g}$ is the density of the gas reservoir and $\rho_{l}^{+}$ is the density of the metastable liquid at the same thermodynamic conditions stabilized by the presence of the planar wall, Eq.
(\ref{eq:ExternalPot_Planar}) and $\delta\approx\frac{1}{2}(\sigma+\sigma_w)$. The off-coexistence of the two phases induces the pressure difference
\begin{align}
p^{+} (\mu) - p(\mu) \approx  \Delta
\rho \Delta \mu, \label{eq:ExpansionPressure}
\end{align}
where $p^{+}$ is the pressure of the metastable liquid and $p$ is
the pressure of the gas reservoir, and where we assume that
$\Delta\mu=\mu-\mu_{sat}<0$ is small.

The excess grand potential per unit area $\mathcal{A}$ of the system
then can be expressed in terms of macroscopic quantities as a
function of $\ell$
\begin{align}
&\frac{\Omega_{ex}(\ell;\mu)}{\mathcal{A}} \label{eq:SIA_GammaEX}\\
 &\qquad = - \Delta \mu \Delta
\rho (\ell - \delta)+ \gamma_{wl}^{\SKA}(\mu) + \gamma_{lg}^{\SKA} + w^{\SKA}(\ell;\mu), \notag
\end{align}
where $\gamma_{wl}^{\SKA}$ and $\gamma_{lg}^{\SKA}$ are the SKA to
the wall-liquid and the liquid-gas surface tensions, respectively,
and $w^{\SKA}(\ell)$ is the effective potential between the two
interfaces (binding potential). In the following, we will suppress
the explicit $\mu$-dependence of these quantities.

The link with the microscopic theory can be made, if the
contributions in the right-hand-side of Eq.~(\ref{eq:SIA_GammaEX})
are expressed in terms of our molecular model, which, when summed
up, give the excess grand potential
(\ref{eq:DefinitionExcessGrandPotential}) where we have substituted
the ansatz (\ref{eq:Sharp-Interface-Profile1D}):
\begin{align}
\gamma_{wl}^{\SKA} &= - \frac{\rho_{l}^{+2} }{2} \int_{-\infty}^0 \int_0^\infty \Phi_{\Pla}\klamm{|z-z'|}dz' dz +\\
& \qquad + \rho_{l}^+ \int_\delta^\infty V_{\infty}(z) dz \notag\\
& =\frac{3}{4} \pi\varepsilon\sigma^4\rho_{l}^{+2}+ \frac{\pi}{90\delta^8}(\sigma_w^{6}-30\delta^6)\sigma_w^6\rho_w\varepsilon_w\rho_{l}^+.\notag\label{gamma_wl_ska}
\end{align}

\begin{align}
\gamma_{lg}^{\SKA} &= - \frac{\Delta \rho^2 }{2} \int_{-\infty}^0 \int_0^\infty \Phi_{\Pla}\klamm{|z-z'|}dz' dz\notag\\
& \qquad =\frac{3}{4}\pi\varepsilon\sigma^4\Delta\rho^2 \\
 w^{\SKA}(\ell) &= \Delta \rho
\left( \rho_{l}^+ \int_{\ell - \delta}^\infty \int_z^\infty \Phi_{\text{\Pla}}(z') dz' dz - \right.\notag\\
&\qquad\left.
 -\int_\ell^\infty V_{\infty}(z) dz \right) \label{eq:PlanarBindingPotential}\\
&= - \frac{A}{12\pi \ell^{2}}
\klamm{
1 +
\frac{2+ 3 \frac{\delta}{\ell}}{1 - \frac{ \rho_{w}  \varepsilon_{w} \sigma_{w}^{6}  }{ \rho_{l}^{+}  \varepsilon \sigma^{6} }}
\frac{\delta}{\ell}+
 O\klamm{{ ({\delta}/{\ell})^{3}   }      }},\notag
\end{align}
where we considered the distinguished limit $\delta \ll \ell$. $A$
is the Hamaker constant given by:
\begin{align}
A =  4 \pi^{2} \Delta \rho \klamm{ \rho_{l}^+\varepsilon\sigma^{6}- \rho_w\varepsilon_w \sigma_{w}^{6}}. \label{eq:HamakerConstant}
\end{align}
We note that the Hamaker constant is implicitly temperature
dependent and that the attractive contribution of the potential of
the wall enables the Hamaker constant to change its sign. Hence, in
contrast with the adsorption on a hard wall, where the Hamaker
constant is always negative, there may be a temperature below which
its sign is positive (large $\rho_{l}$) and negative above. Clearly,
complete wetting is only possible for $A<0$.

Making use of only the leading-order term in
(\ref{eq:PlanarBindingPotential}) the minimization of
(\ref{eq:SIA_GammaEX}) with respect to $\ell$ gives:
\begin{align}
\Delta \rho \Delta \mu - \frac{A}{6\pi \ell^{3}}
\approx 0. \label{eq:PlanarSharpInterfacePrediction}
\end{align}
Hence, at this level of approximation the equilibrium thickness of
the liquid film is:
\begin{align}
\ell_{eq} \approx \klamm{ \frac{A}{6\pi \Delta \rho \Delta \mu }
}^{1/3}. \label{eq:EqFilmThickness}
\end{align}
When substituted into (\ref{eq:SIA_GammaEX}), the wall-gas surface
tension to leading order reads:
\begin{align}
\gamma_{wg}^{\SKA} &=
 \gamma_{wl}^{\SKA}+ \gamma_{lg}^{\SKA} +
\klamm{-\frac{9A}{16\pi} }^{1/3}  { |\Delta \rho\Delta \mu|}^{2/3}.
\end{align}

\begin{figure}[hbt]
\includegraphics[width=8.5cm]{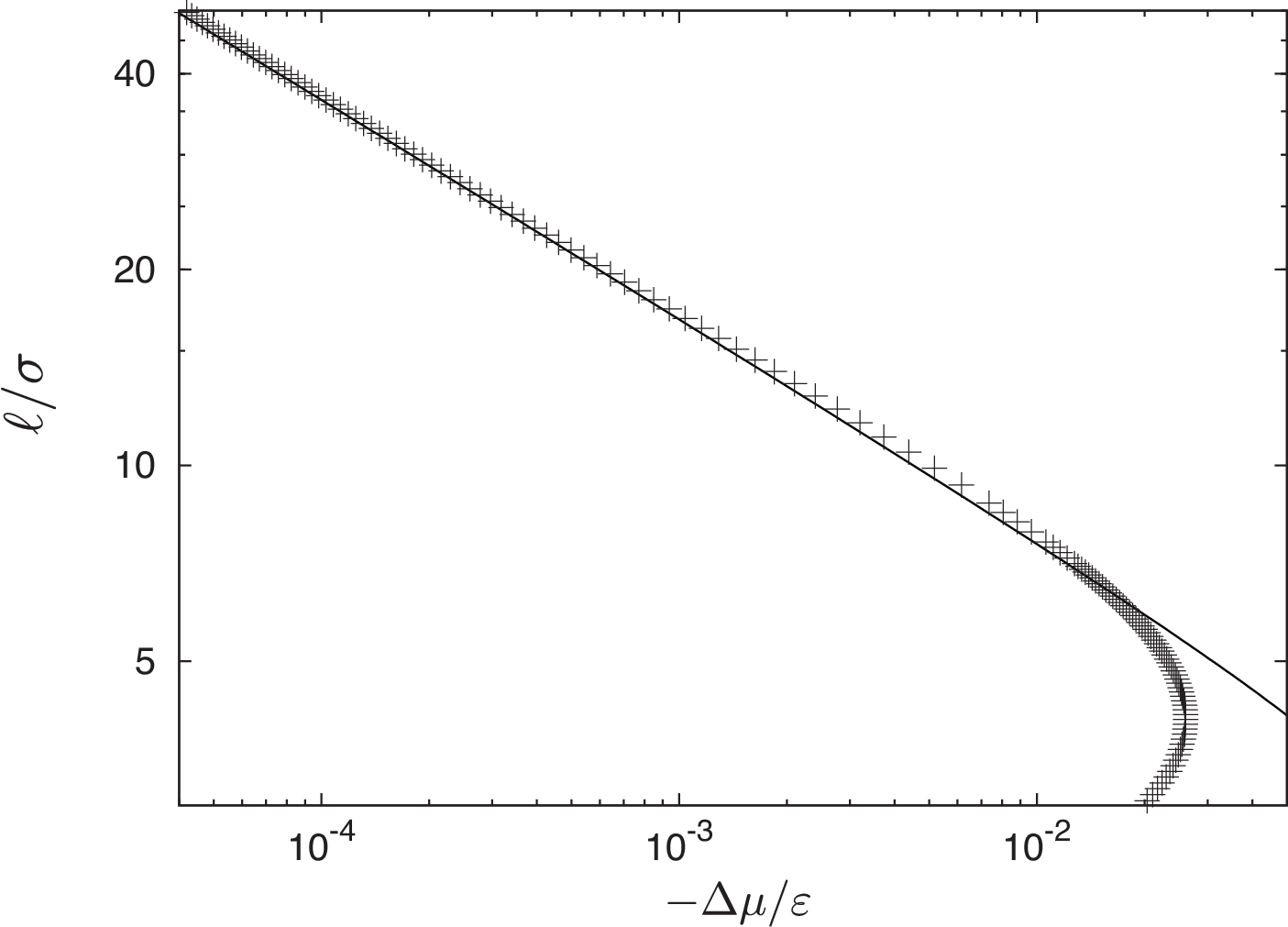}
\caption{Log-log plot of the film thickness as a function of
deviation of the chemical potential from saturation, $\Delta \mu$,
for $k_{B}T= 0.7 \varepsilon$ and wall parameters
$\rho_w\varepsilon_w  = 0.8\varepsilon/\sigma^{3}$, $\sigma_w = 1.25
\sigma$. The crosses are results from DFT computations. The solid
line is the analytical prediction in
Eq.~(\ref{eq:PlanarSharpInterfacePrediction}) obtained from SKA.
\label{fig:SharpInterfacePlanar}}
\end{figure}

Equation~(\ref{eq:EqFilmThickness}) can be confirmed by a comparison
against the numerical DFT, see Fig.~\ref{fig:SharpInterfacePlanar}.
We observe that the prediction of SKA becomes reliable for
$|\Delta\mu|<0.01\varepsilon$ corresponding to a somewhat
surprisingly small value of the liquid film, $\ell\approx5\sigma$.
\newstuff{
Beyond this value, the coarse-grained approach looses its validity
and also the prewetting transition is approached, both of which
cause the curve in Fig.~\ref{fig:SharpInterfacePlanar} to bend (see
also Fig. \ref{fig:PlanarIsotherm}). } It is worth noting that the
only term in (\ref{eq:SIA_GammaEX}) having an $\ell$-dependence and
thus governing the wetting behavior, is the term related to the
undersaturation pressure and the binding potential,
$w^{\SKA}(\ell)$. Clearly, $\gamma_{lg}$ does not come into play in
the planar case since the translation of the liquid-gas interface
along the $z$ axis does not change the free energy of the system.
The situation becomes qualitatively different if the substrate is
curved. Nevertheless, at this stage we conclude in line with
earlier studies, that SKA provides a fully satisfactory approach to
the first-order wetting transition on a planar wall.

\section{Wetting on a curved substrate \label{sec:WettingCurvedSubstrate}}
\subsection{SKA for the spherical wall \label{sec:SKA_curved}}
For the spherical geometry, SKA adopts the following form:
\begin{align}
\rho^{\SKA}_{R,\ell}(r) =  \left\{\begin{array}{ll}
0 &  r<R + \delta,\\
\rho_{l}^{+} & R + \delta<r<R+\ell,\\
\rho_{g} &  R+\ell<r<\infty\, .
\end{array}\right.  \label{eq:SKA_curved}
\end{align}

The corresponding excess grand potential now reads
\begin{align}
\frac{\Omega_{ex}(\mu,R,\ell)}{4\pi R^2}
&= - \Delta \mu \Delta\rho \frac{
\klamm{R+\ell}^{3} - \tilde R^{3}
}{3R^{2}}+ \gamma_{wl}^{\SKA}(R)
+ \notag\\
& +\gamma_{lg}^{\SKA}(R+\ell)\left(1+\frac{\ell}{R}\right)^{2}+w^{\SKA}(\ell; R ), \label{eq:SKA_GammaEX}
 \end{align}
where $\tilde R = R+ \delta$. Within this approximation, the
liquid-vapour surface tension becomes (see also Appendix
\ref{sec:AppendixSKA})

\begin{align}
 \gamma_{lg}^{\SKA}(R)  = \gamma_{lg}^{\SKA}(\infty)
&\left[1 - \frac{2}{9} \frac{\ln (R/\sigma)}{(R/\sigma)^{2}} + O\klamm{ \klamm{{\sigma}/{R}}^{2}   } \right] \label{eq:SKA_GammaLG}
\end{align}
and an analogous expansion holds for $\gamma_{wl}^{\SKA}(R)$.
\newstuff{ 
The $\frac{\ln (R/\sigma)}{(R/\sigma)^{2}}$ correction to $
\gamma_{lg}^{\SKA}(\infty)$ is due to the $r^{-6}$ decay of our
model. We note that short-range potentials lead to different
curvature dependence of the surface tension, a point that has been
discussed in detail in Refs.
\cite{StewartEvans,Evans:2004ys,Parry:2006fk}. Interestingly, the
$O(\sigma/R)$ correction to the surface tension, as one would expect
from the Tolman theory~\cite{Tolman}, is missing. It corresponds to
a vanishing Tolman length within SKA, as we will explicitly show in
the following section.}
 Although the
value of the Tolman length is still a subject of some controversy, it is
most likely that its value is non-zero, unless the system is
symmetric under interchange between the two coexisting
phases~\cite{Fisher:1984kx}. This observation has been confirmed
numerically in Ref.~\cite{StewartEvans} from a fit of DFT results
for the wall-gas surface tension in a non-drying regime for the
hard-wall substrate. Thus, the linear term was included by hand into
the expansion (\ref{eq:SKA_GammaLG})~\cite{StewartEvans}.

Finally, the binding potential within the SKA for the spherical wall yields
\begin{align}
w^{\SKA}\klamm{\ell;R} =w^{\SKA}\klamm{\ell;\infty}\left(1+\frac{\ell}{R}\right)
\end{align}
where terms $O\klamm{(\delta/\ell)^{3},\delta/R,{\frac{\ln
(\ell/R)}{(R/\ell)^{2}}}}$ have been neglected.
\subsection{SIA for the spherical wall \label{sec:SIA_curved}}
\begin{figure}[hbt]
\includegraphics[width=8.5cm]{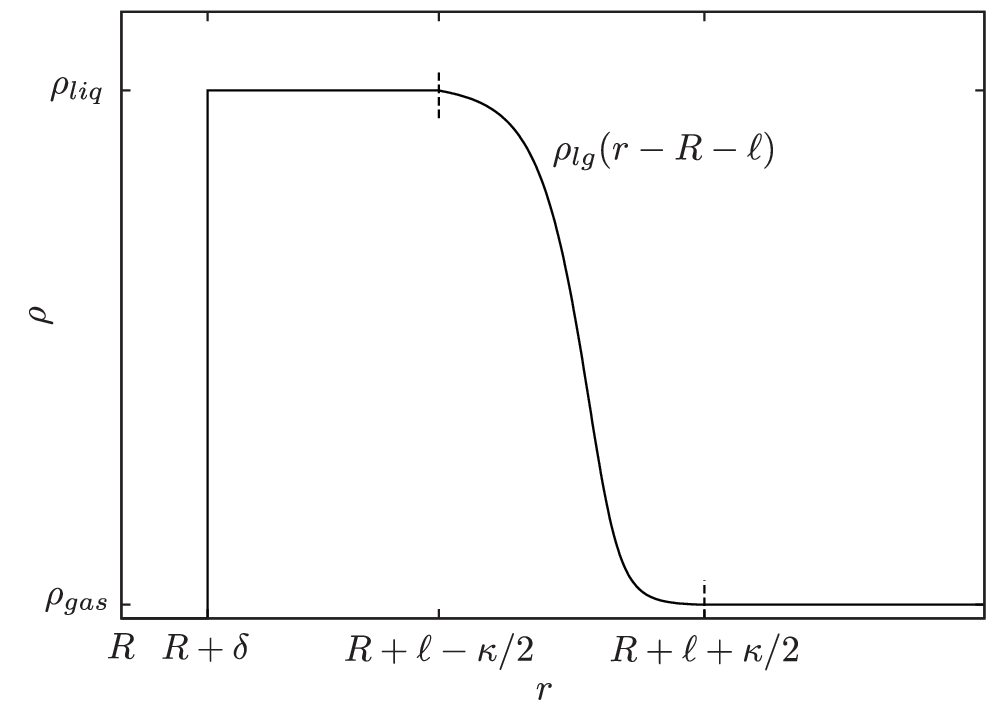}
\caption{Sketch of the density profile according to SIA for a
certain film thickness $\ell$. A piecewise function approximation is
employed so that except for the interval
$(R+\ell-\chi/2,R+\ell+\chi/2)$ the density is assumed to be
piecewise constant.} \label{fig:TestFunction}
\end{figure}

As an alternative to SKA, Napi{\'o}rkowski and
Dietrich~\cite{DietrichNapiorkowski_BulkCorrelation} proposed a
modified version of the effective Hamiltonian, in which the
liquid-gas interface was approximated in a less crude way by a
continuous monotonic function, the SIA. Applied for the second-order
wetting transition on a planar wall, SIA merely confirmed that SKA
provides a reliable prediction for such a system. Formulated now for the spherical case, the
density profile of the fluid takes the form:
\begin{align}
&\rho^{\SIA}_{R,\ell}(r) = \label{eq:SphericalSoftInterfaceApproximation} \\
&=\left\{
\begin{array}{lll}
0 & &  r < R + \delta\\
\rho_{l}^{+} & &   R+\delta <r<R+\ell-\frac{\chi}{2}\\
\rho_{lg}(r-R-\ell) & &  R+\ell-\frac{\chi}{2}<r<R+\ell+\frac{\chi}{2}\\
\rho_{g} & &  R+\ell+\frac{\chi}{2}<r<\infty
\end{array}
\right. . \notag
\end{align}
Thus, a non-zero width of the liquid-vapour interface, $\chi$, is
introduced as an additional parameter. The density profile
$\rho_{lg}(\cdot)$ in this region is not specified, but the
following constraints are imposed:
\begin{align}
\rho_{lg}\klamm{-\frac{\chi}{2}}=\rho_{l}^+ \qquad \text{ and } \qquad \rho_{lg}\klamm{\frac{\chi}{2}}=\rho_{g}, \label{eq:ContinuityCondition}
\end{align}
with an additional assumption of a monotonic behaviour of the
function $\rho_{lg}(r)$. An illustrative example of
$\rho^{\SIA}_{R,\ell}(r)$ is given in Fig.~\ref{fig:TestFunction}.
The corresponding excess grand potential takes the form
\begin{align}
\frac{\Omega_{ex}}{4\pi R^2}
 =&  - \Delta \mu\Delta\rho
\frac{ \klamm{R+\ell}^{3} - \tilde R^{3}
}{3R^{2}}
 +\gamma_{wl}^{\SIA}(R)+ \notag\\
&+
\klamm{1+\frac{\ell}{R}}^{2}
\gamma^{\SIA}_{lg}(R + \ell) + w^{\SIA}\klamm{R,\ell}, \label{eq:OmegaExcessSpherical_FiniteInterface}
\end{align}
taking $R+\ell$ as the Gibbs dividing surface
\newstuff{(so that $\ell$ is a measure of the number of
particles adsorbed at the wall).} The binding potential
(see also Appendix \ref{sec:AppendixSIABindingPot}) is obtained from
\begin{align}
w^{\SIA}&\klamm{R,\ell} =\notag\\
=  &\rho_{l}^+ \int_{R+\ell-\chi/2}^\infty \klamm{\rho_{l}^+ - \rho^{\SIA}_{R,\ell}(r) }  \Psi_{R+\delta}\klamm{r} \klamm{\frac{r}{R}}^2 dr -\notag\\
-&  \int_{R+\ell-\chi/2}^{\infty} \klamm{ \rho_{l}^+
-\rho^{\SIA}_{R,\ell}(r) } V_{R}(r) \klamm{\frac{r}{R}}^2 dr,
\label{eq:SphericalBindingPotential}
\end{align}
where $\Psi_{R}( r)=\int_0^{ R} \Phi_{\Sph}( r, r') d r'$ -- see
Appendix \ref{sex:AppendisSKASurfaceTension} for the explicit form
of the last expression.

The wall-liquid surface tension remains unchanged compared to that
obtained from SKA, Eq.~(\ref{gamma_wl_ska}). However, the liquid-gas
surface tension now reads (see Appendix
\ref{sex:AppendixSIASurfaceTension})
 \begin{align}
 \gamma_{lg}^{\SIA}(R)&= - \int_{R-\chi/2}^{R+\chi/2} \klamm{ p( \rho_{lg,R}( r))-  p_{\text{ref}}}  \klamm{\frac{r}{R}}^2 d r + \notag\\
&+ \frac{1}{2} \int_0^\infty \int_0^\infty \rho_{lg,R}( r) \klamm{\rho_{lg,R}( r') -\rho_{lg,R}( r)}\times \notag\\
& \qquad \times \Phi_{\Sph}( r, r')  \klamm{\frac{r}{R}}^2 d r' d r, \label{gamma_lg_sia}
\end{align}
where $p_{\text{ref}}$ is the pressure at saturation.

\newstuff{
From now on, we neglect the curvature dependence of $\chi$ and
$\rho_{lg,R}\klamm{\cdot}$, as they would introduce higher-order
corrections not affecting the asymptotic results at our level of
approximation. This is also in line with previous studies which show
that the Tolman length only depends on the density profile in the
planar limit \cite{Fisher:1984kx}. Then (\ref{gamma_lg_sia}) can be
written as}
\begin{align}
 \gamma_{lg}^{\SIA}(R)&=\gamma_{lg}^{\SIA}(\infty)\left[1-\frac{2\delta_{\infty}}{R}+{\cal O}\left(\frac{\ln (R/\sigma)}{(R/\sigma)^{2}}\right) \right]\,,
\end{align}
where $\delta_{\infty}$ is the Tolman length of the liquid-gas surface
tension, as given by (Appendix \ref{sec:AppendixSIATolman}):
 \bb
\delta_{\infty} = \frac{1}{\gamma_{lg}^{\SIA}(\infty)} \int_{-\chi/2}^{\chi/2} \klamm{ p(\rho_{lg}(z)) - p_{ref}} z dz. \label{eq:TolmanLengthSIA}
 \ee
The Tolman length is independent of the choice of the dividing surface.
We also note that an
immediate consequence of Eq.~(\ref{eq:TolmanLengthSIA}) is that
within SKA the Tolman length vanishes.

The equilibrium film thickness then follows from setting the derivative
of (\ref{eq:OmegaExcessSpherical_FiniteInterface})
w.r.t. $\ell$ equal to zero:
\begin{align}
& \frac{1}{4\pi R^2}\frac{d\Omega_{ex}}{d \ell}=  - \Delta \mu \Delta \rho \klamm{ 1  + \frac{ \ell}{R}}^2+
\label{eq:DerivativeOmegaExcessFilmThicknessSpherical}\\
&+ \frac{2}{R}\klamm{1+ \frac{\ell}{R}} \gamma_{lg}^{\SIA}( R+ \ell) +  \klamm{ 1+ \frac{\ell}{R}}^2 \left.\frac{d\gamma_{lg}^{\SIA}}{d \ell}\right|_{R+\ell}
+\notag\\
& +  \rho_{l}^+ \int_{ R +  \ell-\chi/2}^{ R +
\ell+\chi/2}  \rho_{lg} '( r- R- \ell) \Psi_{R+\delta}( r)\klamm{\frac{r}{R}}^{2} dr -\notag\\
&- \int_{ R +  \ell-\chi/2}^{ R + \ell+\chi/2}  \rho_{lg} '(
r- R- \ell) V_{R}( r)  \klamm{\frac{r}{R}}^{2} d r. \notag
\end{align}
The last two terms of
(\ref{eq:DerivativeOmegaExcessFilmThicknessSpherical}) are of the
form
\begin{align}
\int_{-\chi/2}^{\chi/2} \rho_{lg}'( r) f_{I,II}( R +  \ell+ r) d r,
\end{align}
with $f_I(r)=\rho_{l}^+ \Psi_{R+\delta}( r) \klamm{\frac{r}{R}}^2$
and $f_{II}(r)= V_{R}( r) \klamm{\frac{r}{R}}^2$. Since
$\rho_{lg}(r)$ is monotonic, i.e. $\rho_{lg}'$ does not change sign,
the mean value theorem can be employed such that
\begin{align}
\int_{-\chi/2}^{\chi/2} \rho_{lg}'( r)
f_{I,II}( R +  \ell&+ r) d r
=\label{eq:PFA_Results1}\\
&- \Delta \rho f_{I,II}( R +  \ell + \xi_{I,II}),\notag
\end{align}
for some  $\xi_{I,II} \in (-\chi/2,\chi/2)$, where we made use
of $\int  \rho_{lg}'( r)  d r = - \Delta  \rho$.
Substituting (\ref{eq:PFA_Results1}) into
(\ref{eq:DerivativeOmegaExcessFilmThicknessSpherical}) and setting
the resulting expression equal to zero, we obtain:
\begin{align}
\Delta \mu =&  \frac{1}{\Delta \rho} \klamm{ \frac{2\gamma_{lg}^{\SIA}(R+\ell)}{R + \ell}
+ \left. \frac{d\gamma_{lg}^{\SIA}}{d\ell}\right|_{R+\ell}   }   - \notag\\
& - \rho_{l}^+ \Psi_{R+\delta}(R + \ell + \xi_I)
\klamm{1+\frac{\xi_I}{R+\ell}}^2+ \notag\\
&+ V_{R}(R + \ell + \xi_{II})\klamm{1+\frac{\xi_{II}}{R+\ell}}^2. \label{eq:WettingCurvedSubstrates_Exact}
\end{align}

So far, there is no approximation within SIA. Equation
(\ref{eq:WettingCurvedSubstrates_Exact}) can be simplified by
appropriately estimating the values of the auxiliary parameters
$\xi_I$ and $\xi_{II}$. To this end, we employ a simple linear
approximation to the density profile at the liquid-gas interface,
taking $- \rho_{lg}'( r)/\Delta \rho \approx 1/\chi$ in
(\ref{eq:PFA_Results1}). Furthermore, we expand $f_{I,II}$ in powers
of $\ell/R, \sigma/\ell$
\begin{align}
f_{I}\klamm{R+\ell + r} = &
-\frac{2\pi \rho_{l}^+\varepsilon\sigma^6}{3\klamm{\ell + r - \delta}^{3}}  \left( 1 +
 \frac{{\ell} + r + 3 \delta }{2R}
+ \right. \notag\\
&\quad
\left.
+O\klamm{
\klamm{\frac{\sigma}{\ell}}^{6},
\klamm{\frac{\ell}{R}}^{2}}
\right), \label{eq:ExpansionfI}\\
f_{II}\klamm{R+\ell + r} = &
-\frac{2\pi \rho_{w} \varepsilon_{w}\sigma_{w}^{6}}{3\klamm{\ell + r}^{3}}  \left( 1 + \frac{{\ell +r }}{2R}+ \right. \label{eq:ExpansionfII}\\
&\qquad \qquad
\left.+
O\klamm{
\klamm{\frac{\sigma}{\ell}}^{6},
\klamm{\frac{\ell}{R}}^{2}
}
\right), \notag
\end{align}
where we assumed the distinguished limits $r,\delta, \sigma \ll \ell
\ll R$. Inserting (\ref{eq:ExpansionfI}) and (\ref{eq:ExpansionfII})
into (\ref{eq:PFA_Results1}) yields for $\xi_{i}$:
\begin{align}
\xi_i =
- \frac{\chi^{2}}{6\ell}
\klamm{ 1+ O\klamm{
\frac{\delta}{\ell}
,\frac{\ell}{R},
\klamm{
\frac{\chi}{\ell}
}^{2}}
}.
\end{align}
From (\ref{eq:WettingCurvedSubstrates_Exact}), we obtain to leading
order,
\begin{align}
\rho_{l}^+\Psi_{R+\delta}&(R + \ell + \xi_I) \klamm{1+\frac{\xi_I}{R+\ell}}^2 = \label{psi}\\
&\qquad \qquad - \frac{2\pi}{3\ell^{3}} \rho_{l}^+\varepsilon\sigma^6
\klamm{1
+O\klamm{
\frac{\delta}{\ell},
\frac{\ell}{R},
\klamm{\frac{\chi}{\ell}}^{2}  }    }
, \notag\\
V_{R}&(R + \ell + \xi_{II})\klamm{1+\frac{\xi_{II}}{R+\ell}}^2 = \label{vr}\\
&\qquad \qquad - \frac{2\pi}{3\ell^{3}} \rho_w\varepsilon_w  \sigma_{w}^{6}
\klamm{1
+O\klamm{
\frac{\ell}{R},
\klamm{\frac{\chi}{\ell}}^{2}  }    }. \notag
\end{align}
Finally, substituting (\ref{psi}) and (\ref{vr}) into
(\ref{eq:WettingCurvedSubstrates_Exact}) we have to leading order:
\begin{align}
\Delta \rho \Delta \mu  - \frac{2}{R}\gamma_{lg}^{\SIA}(\infty)
 \approx &  \frac{A}{6\pi \ell^{3}}, 
 \label{eq:LargeRSphericalAsymp}
\end{align}
and hence, to leading order the equilibrium wetting film thickness
is:
\begin{align}
  \ell_{eq}^{\SIA} \approx \klamm{ \frac{A}{6\pi \klamm{ \Delta \rho \Delta \mu - 2 \gamma_{lg,\infty}^{\SIA}/R }   } }^{1/3}.
  \label{l_sia}
\end{align}

We note that this asymptotic analysis can be extended beyond
(\ref{l_sia}), by including terms $O(\delta/\ell)$, $O(\ell/R)$ and
$O((\chi/\ell)^{2})$. The latter occurs due to the ``soft"
treatment of the liquid-vapor interface and is thus not present in
SKA.

\begin{figure}[hbt]
\includegraphics[width=8.5cm]{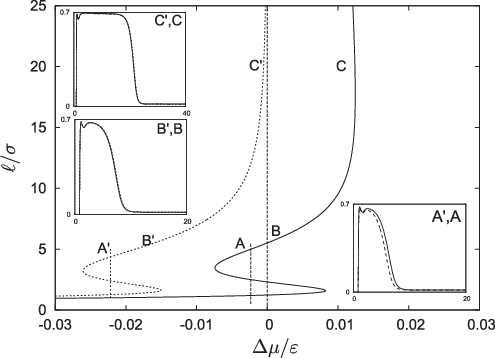}
\caption{Isotherms and density profiles for a planar wall (dashed
lines) and a sphere with $R = 100\sigma$ (solid lines) at $k_BT= 0.7
\varepsilon $ and with wall parameters, $\rho_w\varepsilon_w =
0.8\varepsilon/\sigma^{3}$ and $\sigma_w = 1.25\sigma$. To directly
compare the planar to the spherical case, the film thickness instead
of adsorption is used as a measure. The subplots in the inset depict
the density $\rho\sigma^{3}$ as a function of the distance from the
wall $z/\sigma$ and $(r-R)/\sigma$ for the planar and the spherical
cases, respectively. The points $A$ and $A'$ are at the prewetting
transitions. Points $B,B'$ and $C,C'$ correspond to the same film
thickness. $B$ is at saturation whereas $C$ is chosen such that the
film thickness $\ell$ is $20\sigma$.
\label{fig:SphericalIsothermProfiles} }
\end{figure}

\begin{figure}[htbp]
\begin{center}
\includegraphics[width=8.5cm]{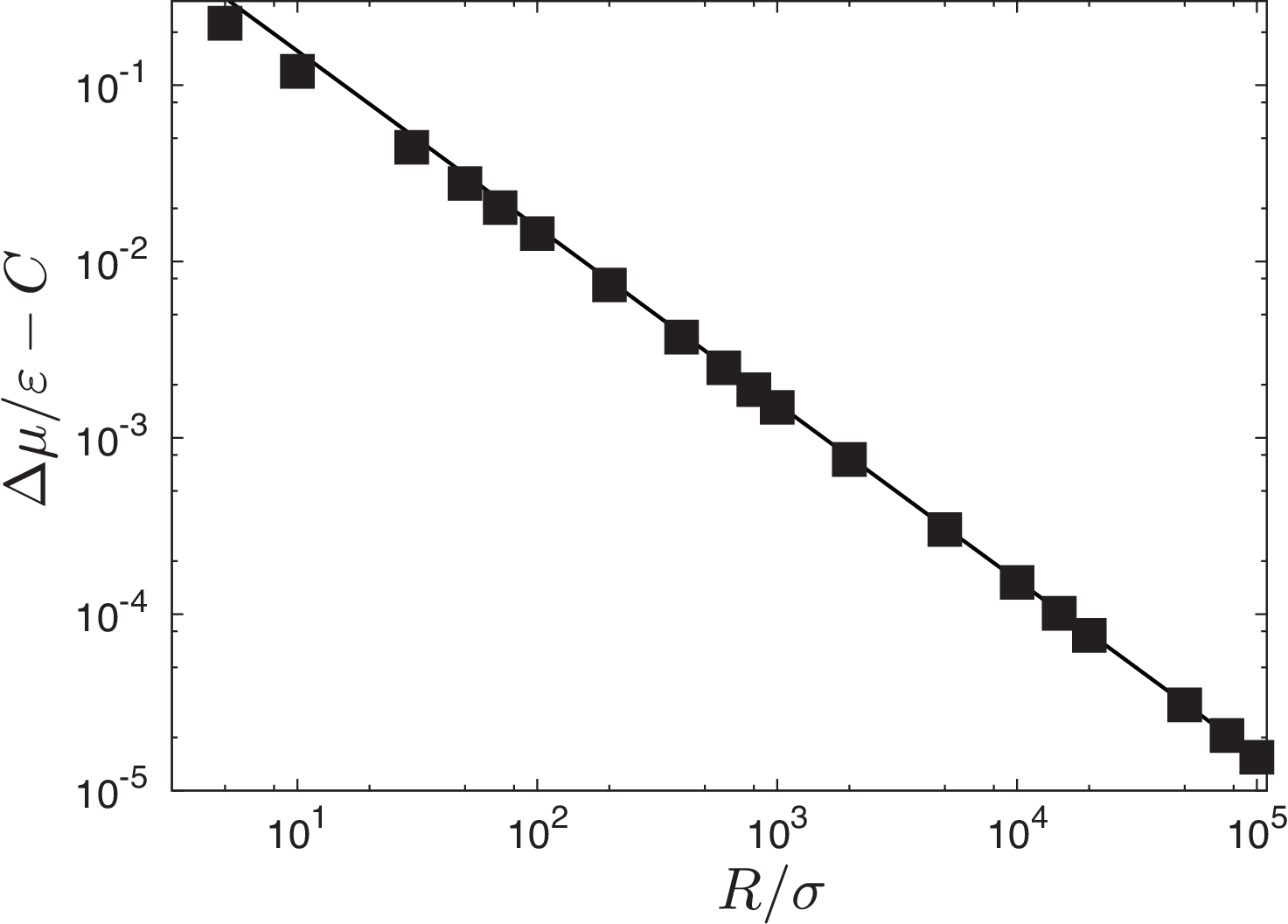}
\caption{Numerical verification of
Eq.~(\ref{eq:LargeRSphericalAsymp}). The film thickness $\ell$ is
fixed and corresponds to the adsorption $\Gamma_{R} =
3.905/\sigma^{2}$. The solid line corresponds to the analytical
result, $\Delta \mu -  2{\gamma_{lg}^{\SIA}(\infty)}/\klamm{\Delta
\rho R} = C\varepsilon$, where $\gamma_{lg}^{\SIA}(\infty) =
0.524\varepsilon/\sigma^{2}$, see Table~\ref{tab:TestFunctions}. The
symbols denote the numerical DFT results.
} \label{fig:DmuOverR}
\end{center}
\end{figure}

In Fig.~\ref{fig:SphericalIsothermProfiles} we compare two
adsorption isotherms ($k_BT=0.7 \varepsilon $) corresponding to
wetting on a planar and a spherical wall ($R=100\sigma$). The two
curves are mutually horizontally shifted by a practically constant
value, in accordance with Eq.~(\ref{eq:LargeRSphericalAsymp}). This
implies that the curve for the spherical wall crosses the saturation
line $\Delta\mu=0$ at a finite value of $\ell$, and eventually
converges to the saturation line as $\Delta\mu^{-1}$ from the right,
thus the finite curvature prevents complete wetting. The horizontal
shift corresponds to the Laplace pressure contribution,
$\Delta\mu={2\gamma_{lg}^{\SIA}(\infty)}/\klamm{\Delta\rho R}$, as
verified by comparison with the numerical DFT,
Fig.~\ref{fig:DmuOverR}. All these conclusions are in line with SKA.
However, the difference between SKA and SIA consists in a different
treatment of $\gamma_{lg}(\infty)$, compare
(\ref{eq:LiqGasSurfaceTensionSKA}) and
(\ref{eq:SIASurfaceTensionPlanar}). This is quite obvious, since the
softness of the interface influences the free energy required to
increase the film thickness. We will discuss this point in more
detail in the following section.

\subsection{Comparison of SKA and SIA \label{sec:SIA_solution}}

We now examine the repercussions of the way the liquid-gas interface
is treated on the prediction of wetting behaviour on a spherical
surface.
As already mentioned in Sec.~\ref{sec:SIA_curved}, the linear
correction in the curvature to the planar liquid-gas surface
tension, ignored within SKA, is properly captured by SIA.
Furthermore, the presence of the Laplace pressure suggests that the
liquid-gas surface tension plays a strong part in the determination
of the equilibrium film thickness. This contrasts to the case of a
planar geometry, where the term associated with the liquid-gas
surface tension has no impact on the equilibrium configuration.

To investigate this point in detail, we will first compare the
approximations of $\gamma_{lg}$ as obtained by the two approaches.
For this purpose, we start with SIA for a given parameterization of
the liquid-gas interface. As shown in Table \ref{tab:TestFunctions},
we employ linear, cubic and hyperbolic tangent auxiliary functions,
where the latter violates condition (\ref{eq:ContinuityCondition})
negligibly. The particular parameters are determined by minimization
of a given function with respect to the corresponding parameters. In
Table \ref{tab:TestFunctions} we display the planar liquid-gas
surface tension associated with a particular parameterization and
the Tolman length resulting from Eq.~(\ref{eq:TolmanLengthSIA}) for
the temperature $k_BT=0.7\varepsilon $. In all three cases the
surface tension is close to the one obtained from the numerical
solution of DFT and also the predictions of the Tolman length are in
a reasonable agreement with the most recent simulation
\newstuff{ results~\cite{Sampayo,Block:2010vn,Giessen:2009kx}, with thermodynamic results \cite{Bartell:2001vn} as well as
with results from the van der Waals square gradient theory \cite{Blokhuis:2006uq}.}

\begin{table}[htdp]
\caption{Planar surface tensions (\ref{eq:SIASurfaceTensionPlanar}), Tolman lengths (\ref{eq:TolmanLengthSIA}) and the
corresponding parameters for temperature $k_BT=0.7 \varepsilon$
according to a given auxiliary function approximating the density
distribution of the vapour-liquid interface. The parameters are from
auxiliary function minimization. The surface tension given by
numerical DFT computations is $\gamma_{lg}=
0.517\varepsilon/\sigma^2$ and $\bar \rho = (\rho_{l}+\rho_{g})/2$. \newstuff{Note that
in the $\tanh$-case, the interface width is implicitly determined by the steepness
parameter $\alpha$.}}
\begin{center}
\begin{tabular}{lccc} \hline 
 Auxiliary function $\rho_{lg}(z)$ 
& $\gamma_{lg}^{\SIA}(\infty)$
& argument & $\delta_{\infty}$\\ \hline
$ \bar \rho - \Delta \rho \frac{z}{ \chi } $
& $0.544 \varepsilon/\sigma^{2}$ & $\chi = 4.0\sigma$ &$ -0.07 \sigma$\\
$\bar \rho - \frac{3}{2}\Delta \rho \frac{z}{\chi } + 2 \Delta \rho \klamm{\frac{z}{ \chi}}^{3}$
 & $0.532 \varepsilon/\sigma^{2}$ & $\chi =5.4\sigma$ & $-0.09 \sigma$\\
$\bar \rho - \frac{\Delta \rho}{2} \tanh \klamm{ \alpha z/\sigma }$
 & $0.524 \varepsilon/\sigma^{2}$ & $\alpha = 0.66$ & $-0.11 \sigma$\\ \hline
\end{tabular}
\end{center}
\label{tab:TestFunctions}
\end{table}%

\begin{figure}
\includegraphics[width=8.5cm]{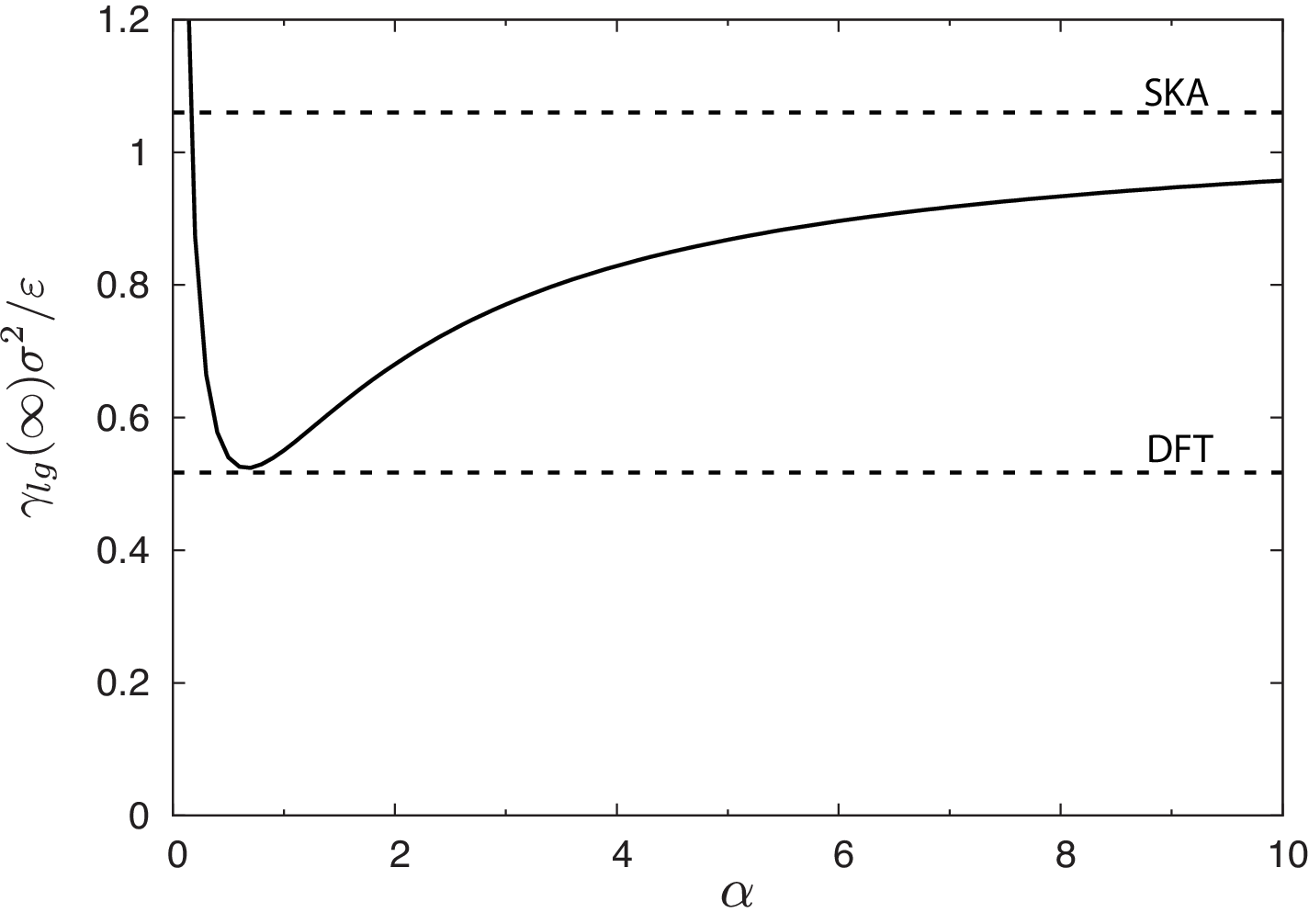}
\caption{Plot of a dimensionless planar liquid-gas surface tension
for the liquid-gas interface approximation $\rho(z)=
\frac{\rho_{l}+\rho_{g}}{2} - \frac{\Delta\rho}{2} \tanh \klamm{
\alpha z /\sigma}$ for $k_{B}T=0.7\varepsilon$ as a function of the
steepness parameter $\alpha$. The upper dashed line is the surface
tension obtained from SKA, whereas the lower dashed line displays
the surface tension obtained from numerical DFT.
}\label{fig:SurfaceTensionSteepness}
\end{figure}

It is reasonable to assume that from the set of considered
auxiliary functions, the $\tanh$-approximation is the most realistic
one, although the numerical results as given in Table
\ref{tab:TestFunctions} suggest that it is mainly the finite width
of the liquid-gas interface, rather than the approximation of the
density profile at this region, that matters. To illustrate this, we
show in Fig.~\ref{fig:SurfaceTensionSteepness} the dependence of the
surface tension on the steepness parameter $\alpha$, determining the
shape of the $\tanh$ function. Note that the limit $\alpha \to
\infty$ corresponds to the surface tension as predicted by SKA,
$\gamma_{lg,\infty}^{\SKA} = 1.060\varepsilon/\sigma^2$, for
$k_{B}T=0.7 \varepsilon $. Such a value contrasts with the result of
SIA, which corresponds to the minimum of the function, and yields
$\gamma_{lg,\infty}^{\SIA} = 0.524\varepsilon/\sigma^2$, in much
better agreement with the numerical solution of DFT,
$\gamma_{lg,\infty}^{\text{DFT}} = 0.517\varepsilon/\sigma^2$.

\begin{figure}[hbt]
\includegraphics[width=8.5cm]{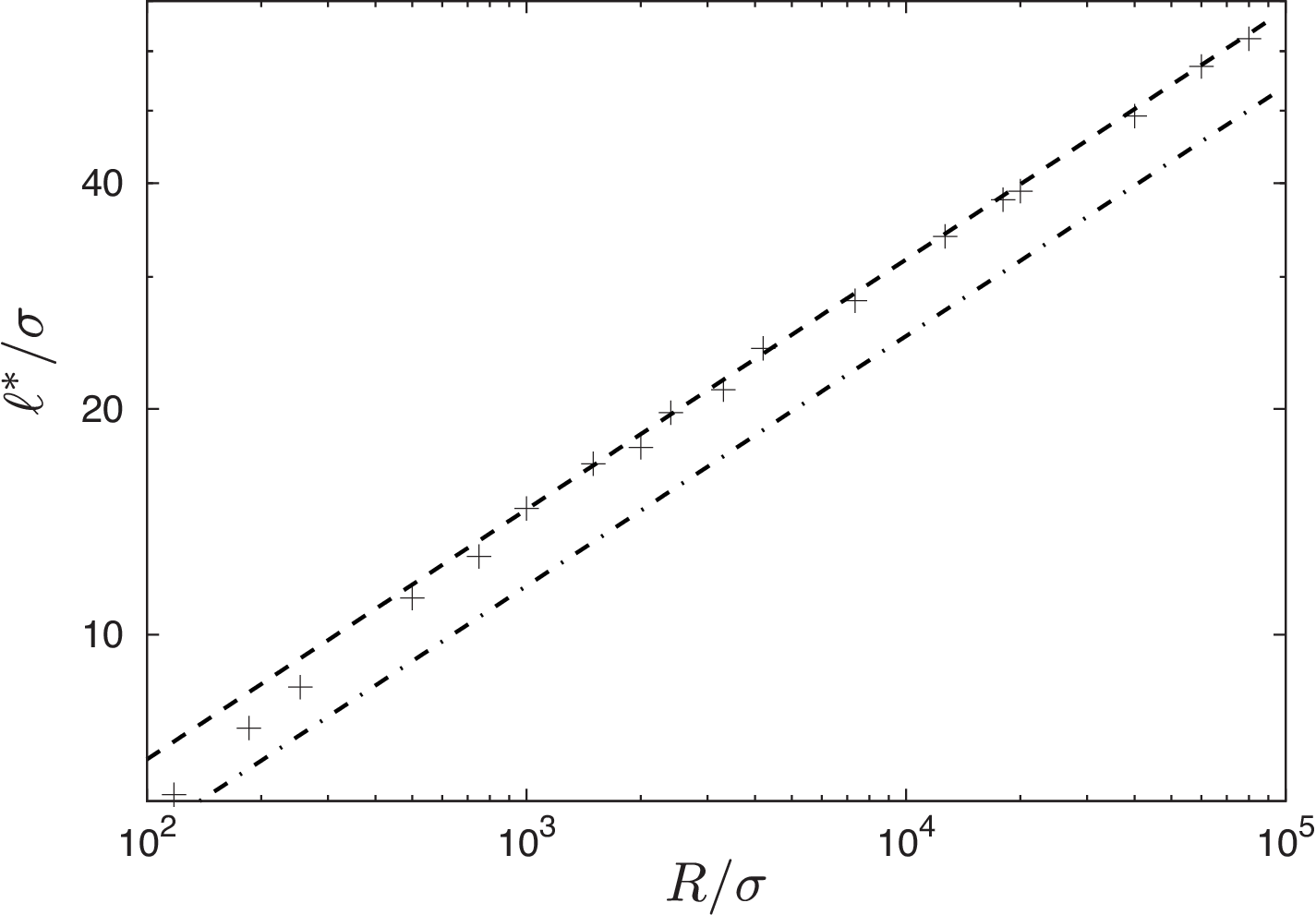}
\caption{Film thickness at saturation ($\Delta \mu = 0$) as a
function of the wall radius. The symbols correspond to the numerical
DFT results. The dashed line shows the prediction according to
Eq.~(\ref{l_sia}), where $\gamma_{lg}^{\SIA}(\infty) = 0.524
\varepsilon/\sigma^{2}$ (see Table~\ref{tab:TestFunctions}). The
dash-dotted line corresponds to Eq.~(\ref{l_sia}) where
$\gamma_{lg}^{\SKA}(\infty) = 1.060 \varepsilon/\sigma^{2}$ is used
instead of $\gamma_{lg}^{\SIA}(\infty)$.
The wall parameters are $\rho_w\varepsilon_w = 0.8\varepsilon/\sigma^{3}$ and $\sigma_w =
1.25\sigma$ at $k_{B}T = 0.7\varepsilon$.}
\label{fig:FilmThicknessAtSaturation}
\end{figure}

Asymptotic analysis of the film thickness in Eq.
(\ref{eq:LargeRSphericalAsymp}), reveals that the film thickness for
large but finite $R$ remains finite even at saturation with
$\ell\sim R^{\frac{1}{3}}$ in line with earlier studies, e.g.
Refs.~\cite{DietrichWettingOnCurvedSubstrates,StewartEvans}. From
Eq.~(\ref{eq:LargeRSphericalAsymp}) one also recognizes a strong
dependence of $\ell$ on the planar liquid-gas surface tension. In
Fig.~\ref{fig:FilmThicknessAtSaturation} we present the SIA and SKA
predictions of the dependence on $\ell$ as a function of the wall
radius. The comparison with the numerical DFT results reveals that
for large $R$ SIA is clearly superior, reflecting a more realistic
estimation of the liquid-gas surface tension.
\newstuff{For small values of $R$ (and $\ell$) we observe a deviation between DFT and the SIA results.
This indicates a limit of validity of our first-order analysis  and the assumption of large film thicknesses.}

\begin{figure}[hbt]
\includegraphics[width=8.5cm]{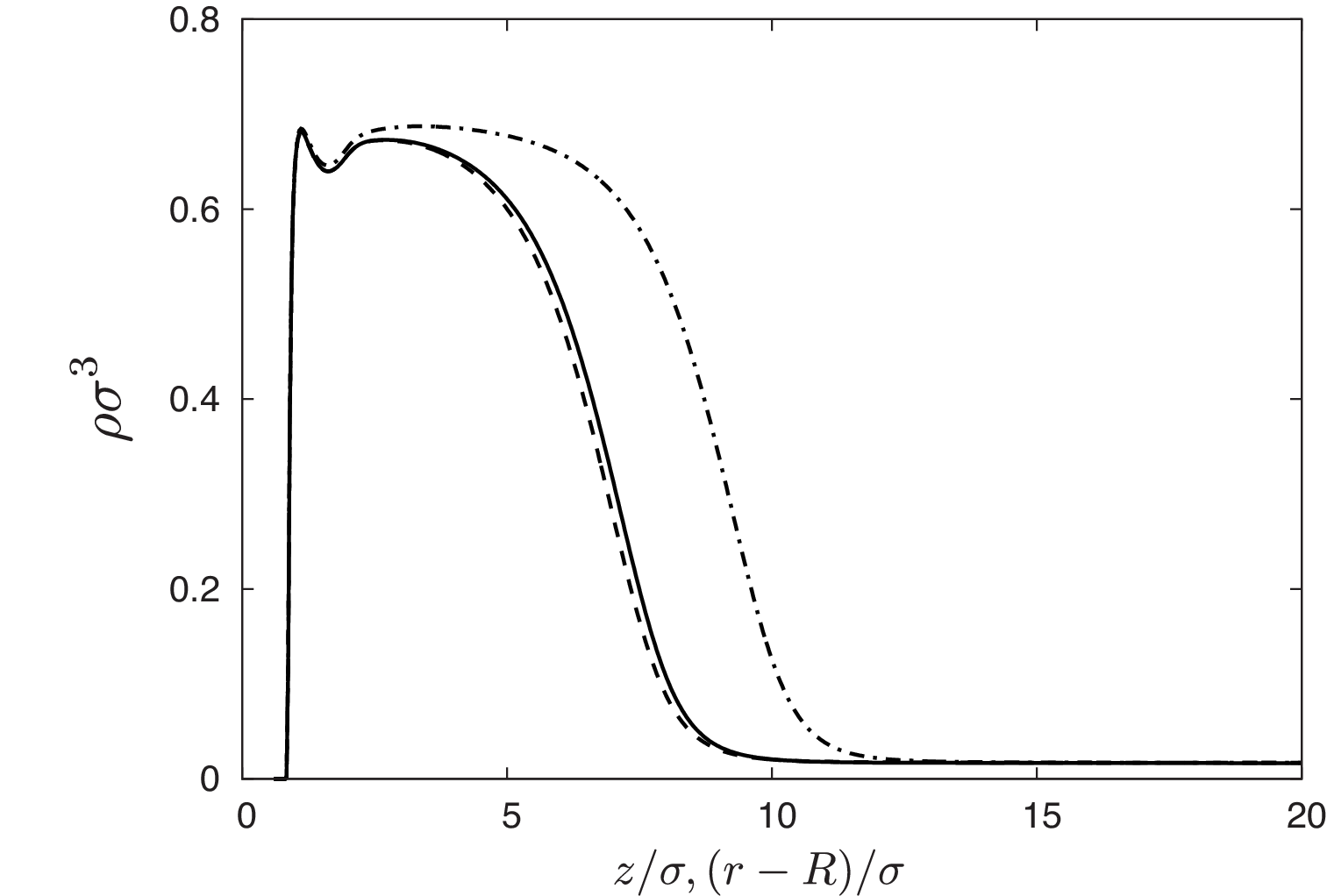}
\caption{ \newstuff{Density profiles of the fluid adsorbed at the spherical walls of radii $R = 104.1\sigma$ (dashed) and $R = 210.6\sigma$ (dashed-dotted)
in a saturated state and at the planar wall (solid line) in an undersaturated state, $\Delta \mu= -0.015\varepsilon$ .
The wall radii correspond to the equality $2 \gamma_{lg,\infty}^j/R=\Delta \rho |\Delta \mu|$ for $j=\rm{SIA}$ (dashed) and $j=\rm{SKA}$ (dashed-dotted).
For $k_{B}T= 0.7\varepsilon$.
The wall
parameters are $\rho_{w}\varepsilon_w = 0.8\varepsilon/\sigma^{3}$ and
$\sigma_w = 1.2\sigma$.}}
\label{fig:ComparePlanarSphericalDensityProfiles}
\end{figure}

The occurrence of the undersaturation pressure and the Laplace
pressure on the left-hand-side of
Eq.~(\ref{eq:LargeRSphericalAsymp}) suggests a certain equivalence
between the two systems of a planar and a spherical symmetry once
the sum of the two pressures is fixed.
In Fig.~\ref{fig:ComparePlanarSphericalDensityProfiles} we test this
equivalence on the level of a density profile, where DFT results
corresponding to the planar and the spherical case are compared,
such that $\Delta\rho|\Delta\mu|=2\gamma_{lg}^{j}(\infty)/R$, with
$j = \{\SIA,\SKA \}$. A high value of $\gamma_{lg}(\infty)$ as given
by SKA must now be compensated by a fairly large $R$. As we have
seen in Fig.~\ref{fig:SphericalIsothermProfiles}, the high value of
$R$ means that the saturation line $\Delta\mu=0$ is crossed by the
adsorption isotherm at large $\ell$, in agreement with the result
depicted in Fig.~\ref{fig:FilmThicknessAtSaturation}. However, for a
{\it given} $R$, $\ell$ as obtained by SKA is underestimated, which
follows from (\ref{l_sia}) with
$\gamma_{lg}(\infty)=\gamma_{lg}^\SKA(\infty)$ which is also
consistent with the physical observation that high surface tension
inhibits growth of the liquid film.

\newstuff{
Note that these results are not in conflict with previous studies
\cite{StewartEvans}, where the SKA has been applied for drying on a
spherical hard wall and very good agreement was obtained with DFT
computations. This is because in Ref. [8] the ``exact'' (i.e.
obtained from DFT computations) liquid-vapor surface tension was
implemented into SKA with a view to verify the correctness of its
functional form. Here, we show that the coarse-grained effective
Hamiltonian approach is capable of a quantitatively reliable
prediction of the adsorption phenomena on a spherical wall (for a
sufficiently large $R$), if the restriction of the sharp liquid-gas
interface is dropped. However, the price we have to pay for, is one
more parameter (compared to SKA) that steps into the theory.}

\section{Summary and conclusions\label{sec:summary}}

We have re-examined the properties of a well known coarse-grained
interfacial Hamiltonian approach, originally proposed by Dietrich
\cite{Dietrich} for the study of wetting phenomena on a planar
substrate and based on a sharp-kink approximation (SKA). SKA relies
on approximating the density profile by a piecewise constant
function and has proved to provide significant insight into
interfacial phenomena as it is mathematically tractable and gives
reliable results for a wide spectrum of problems. This theory is
phenomenological in its origin, but a link with a microscopic
density functional theory (DFT) can be made, which allows to express
all the necessary quantities in terms of fluid-fluid and
fluid-substrate interaction parameters. Comparison with numerical
DFT reveals that SKA provides a fully satisfactory approach to the
theory of complete wetting on a planar surface.

One of the aims of this study was to demonstrate that for a
spherical geometry the prediction quality of SKA regarding
interfacial properties and wetting characteristics is limited. More
specifically, we demonstrated that SKA satisfactorily determines the
functional form of the asymptotic behaviour of the film thickness
for large radii of the substrate but leads to a significant
quantitative disagreement in the prediction of the adsorbed film
thickness when compared against numerical DFT. The source of the
deviation is the presence of the Laplace pressure that is
\newstuff{not quantitatively captured within the framework of SKA.}
This contribution originates in the dependence of the free energy of the
the liquid-gas interface on a position of a dividing surface, a
property that is absent in the planar case.

We then showed that the properties of the effective interfacial
Hamiltonian approach can be substantially improved if SKA is
replaced by a soft-interface approximation (SIA), where the
assumption of the sharp liquid-gas interface is replaced by a less
restrictive approximation in which the interface is treated as a
continuous function of the density distribution. We demonstrated
that SIA allows for mathematical scrutiny as it is still
analytically tractable, e.g. it provides the curvature expansion of
the surface tensions (non-analytic in the wall curvature) with the
leading-order term proportional to $\sigma/R$. Moreover, it allows
to express the corresponding coefficient, the Tolman length, in a
fairly simple manner and the values it predicts for the Tolman
length are in a reasonable agreement with the latest simulation
results.

This is in contrast with SKA, where the linear term in the
surface tension expansion is missing, i.e. the Tolman length vanishes.
 This observation is in a full agreement with the
conclusion of Fisher and Wortis \cite{Fisher:1984kx}, since SKA
treats the fluid in a symmetric way, and thus the Tolman length must
disappear as for the Ising-like models. In other words, according to
SKA, the surface tension of a large drop is equivalent to the one of
a bubble, provided the density profiles of the two systems are
perfectly antisymmetric in the planar limit. This is no more true
for SIA, due to the asymmetry of the ``local" contributions to the
surface tension, i.e. the first term on the right-hand-side of
Eq.~(\ref{gamma_lg_sia}).

\newstuff{Furthermore, comparison with our numerical DFT revealed that the SIA
results of the film thickness as a function of the wall radius are
offer a drastic improvement to the ones obtained from SKA. This
follows from the fact that the surface tension of the planar
liquid-gas interface according to SKA is overestimated, which in
turn underestimates the interface growth.}

\begin{figure}[htbp]
\begin{center}
\includegraphics[width=8.5cm]{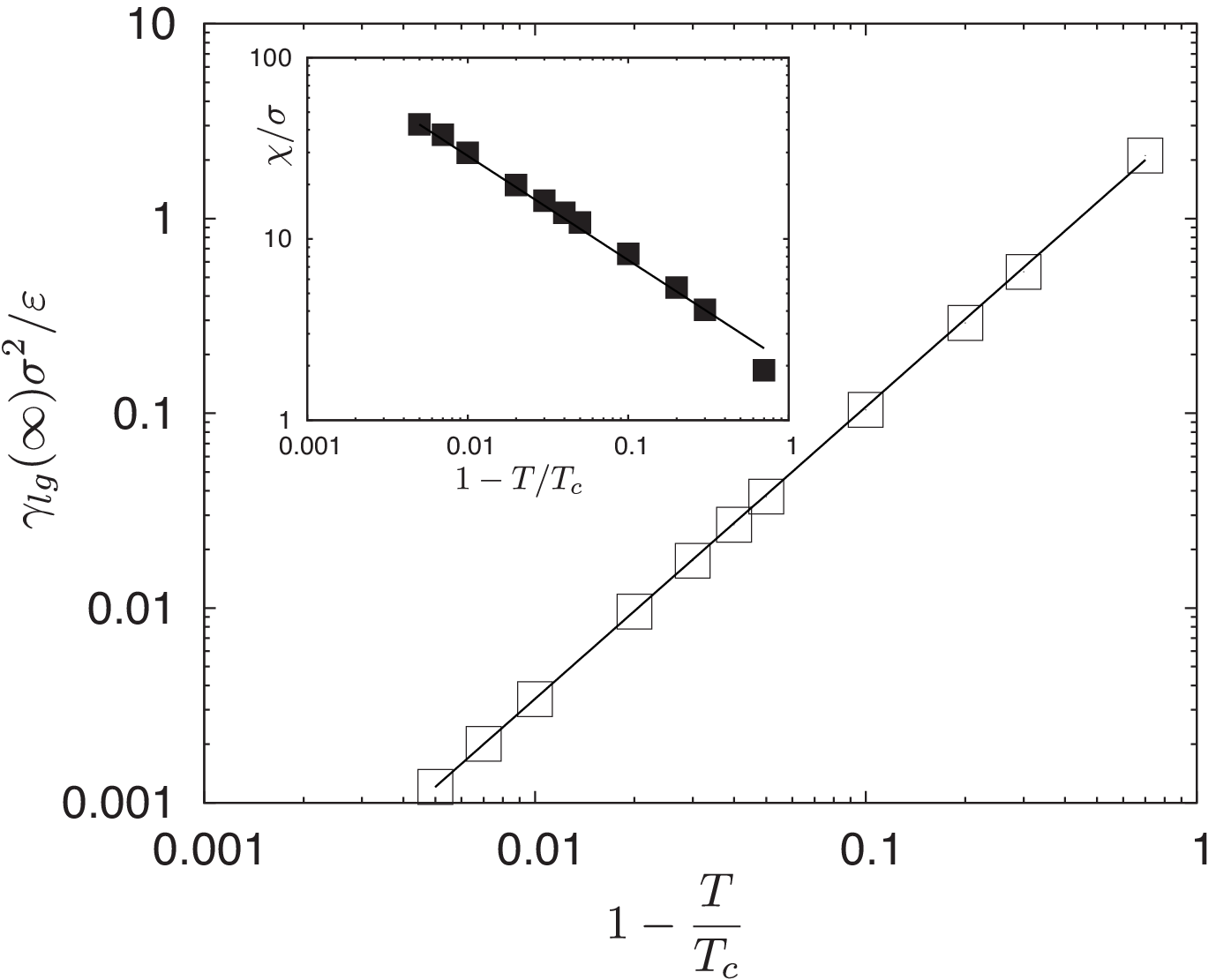}
\caption{\newstuff{Plot of liquid-gas surface tension vs. $t = 1- T/T_{c}$.
The squares are the result of SIA, where a simple linear interpolant has been used
to model the interface density profile. The surface tension
has been obtained by minimizing the grand potential with respect to the interface width $\chi$. The
solid line is a fit to $\gamma_{lg}(\infty)\sigma^{2}/\varepsilon = C t^{3/2}$, where the resulting coefficient is $C = 3.4$.
The inset shows a plot of the interface width $\chi/\sigma$ over $t$. The solid line is a fit to $\chi = C_{\chi}t^{-\alpha}$, where
$C_{\chi} = 2.0$ and $\alpha = 0.57$.}}
\label{fig:CriticalBehavior}
\end{center}
\end{figure}

It should be emphasized that all the theoretical approaches we have
considered in this work are of a mean-field character, i.e. do not
properly take into account the interfacial fluctuations (capillary
waves) at the liquid-gas interface. However, for our fluid model of
a power-law interaction these fluctuations are not expected to play
any significant role, since the upper critical dimension associated
with the considered system is $d_c^*=2$~\cite{Lipowsky:1984uq}. Nevertheless, what one
has to take into account in order to obtain the correct
critical behavior, is the broadening of the interface at
the critical region. Evidently, this feature is not provided
by SKA. Consequently, within SKA the liquid-gas surface
tension vanishes as $t = 1 - \frac{T}{T_{c}}$\cite{DietrichWettingOnCurvedSubstrates}. \newstuff{In contrast, the SIA provides the expected mean-field
behavior $\gamma_{lg}(\infty) \sim t^{\frac{3}{2}}$, as it is able to capture the interface
broadening near the critical point (see Fig. \ref{fig:CriticalBehavior}).}

\newstuff{The SIA developed here, can be naturally extended by ``softening'' the
wall-liquid interface in an analogous way as done for the
liquid-vapor interface. However, such a modification would have
presumably only negligible impact on the prediction of the thickness
of the adsorbed liquid film, since the contribution to the excess
free energy from the wall-liquid surface tension has no
$\ell$-dependence and the change of the binding potential is
expected to be small. On the other hand, it may be interesting to
find the influence of this refinement on quantities such as the
density profile at contact with the wall. For this purpose a
non-local DFT (e.g. Rosenfeld's fundamental measure theory) would be
needed though~\cite{StewartEvans,Stewart:2005fk,Blokhuis:2007uq}.}

We also note that despite our restriction to a model of spherical symmetry,
our conclusions should be relevant for general curved geometries and
should capture some of the qualitative aspects of wetting on
non-planar substrates. Of particular interest would be the extension
of this study to spatially heterogeneous, chemical or topographical
substrates. Such substrates have a significant effect on the wetting
characteristics of the solid-liquid pair (e.g.
Refs.~\cite{JCollInt_202_1998,PhysFluids_16_2004,Que07,PhysFluids_21_2009,PhysRevLett_104_2010,Bohlen2008}).

\acknowledgements{\newstuff{We are grateful to Bob Evans for
valuable comments and suggestions on an early version of the
manuscript and for bringing to our attention Ref.
\cite{Stewart:2005fk}. We thank} Antonio Pereira for helpful
discussions regarding numerical aspects of our work. This work is
supported by the Rotary-Clubs Darmstadt, Darmstadt-Bergstra\ss e and
Darmstadt-Kranichstein, Engineering and Physical Sciences Research
Council of England Platform Grant No. EP/E046029 and EU-FP7 ITN
Multiflow and ERC Advanced Grant No. 247031. AM is grateful to the
financial support of the Ministry of Education, Youth and Sports of
the Czech Republic under Project No. LC512 and the GAAS of the Czech
Republic (Grant No. IAA400720710).}

\appendix

\section{Numerical methods \label{sec:Annay:NumericalMethod}}
For our computations we employ dimensionless values. We use $\sigma$
and $\varepsilon$, as the characteristic length and energy scales,
respectively.

\subsection{Density profile \label{sec:Annex:DensityProfiles}}

To obtain the equilibrium density profiles, the extremal conditions
Eqs.~(\ref{eq:ExtremalCondition1D})
and~(\ref{eq:ExtremalConditionSph}) for the planar and the spherical
case, respectively, must be solved numerically. As both cases are of
dimension one, the same numerical method can be applied and we
restrict ourselves to presenting the numerical method for the planar
wall, $W = \mathbb{R}^2\times \mathbb{R}^-$.

The domain $\mathbb{R}$ normal to the wall is restricted to an interval of interest $[
z_0, z_N]$ with boundary conditions $\rho( z) = 0$ for $ z <  z_0$
and $\rho( z) = \rho_{g}$ for $ z > z_N$. $z_0 \in (0,1)$ is
typically chosen to be $0.6$. This can be done due to the repulsive
character of the wall. The interval $[ z_0, z_N]$ is then divided in
a uniform mesh, $ z_i = z_0 + i\cdot \Delta  z$ with $i = 0,\ldots
N$, where $\Delta z = ( z_N -  z_0)/N$ is the grid size.
Subsequently, the integral in Eq.~(\ref{eq:ExtremalCondition1D}) is
discretized using a trapezoidal rule inside the domain $[ z_0,
z_N]$, whereas the analytical expression
\begin{align*}
\Psi_{\Pla}\klamm{z} &= \int_{z}^{\infty} \Phi_{\Pla}(z') dz'\\
&= \left\{
\begin{array}{ll}
\klamm{- \frac{16}{9} \pi + \frac{6}{5} \pi \frac{z}{\sigma}}\varepsilon \sigma^{3} & \text{if } z < \sigma\\
4\pi\varepsilon \sigma^{3} \klamm{\frac{1}{45}\klamm{\frac{\sigma}{z}}^{9}  - \frac{1}{6}\klamm{\frac{\sigma}{z}}^{3} }
& \text{if } z \geq \sigma
\end{array}
\right.
\end{align*}
 is used for the
integral outside that interval. Hence, we obtain a system of $N+1$
nonlinear equations with $\{\rho_i, i = 0,\ldots,N\}$ as unknowns,
namely:
\begin{align}
g_i&(\rho_0,\ldots,\rho_{N}) \defi \mu_{\HS}\klamm{ \rho_i}+  V_{\infty}( z_i) - \mu + \label{eq:DiscretizedEquation} \\
& + \rho_{g} \Psi_{\Pla}\klamm{z_N -  z_i} +\notag\\
&+\frac{\Delta  z}{2} \sum_{j=1}^{N-1} \klamm{2 - \delta_{j0} -\delta_{jN}}\rho_j \Phi_{\Pla}\klamm{| z_j- z_i|} \istobe
0,\notag
\end{align}
where $\delta_{ij}$ denotes the Kroenecker-Delta, which we have
used in order to take into account the grid size at the boundaries.

This system of equations is solved using a modified Newton method,
where {each step ${\Delta} {\mbox {\boldmath $ \rho$}}$ is
rescaled with a parameter $\lambda$ such that ${\mbox{\boldmath $
\rho$}}^{n+1} = {\mbox{\boldmath$\rho$}}^n + \lambda {\Delta} {\mbox
{\boldmath $ \rho$}}$ is bounded in $(0,6/\pi)$} in order to avoid
the singularity of Eq.~(\ref{eq:DefinitionPressure}). Note that we
have made use of the vector notation ${\mbox{\boldmath $ \rho$}}
\defi \klamm{\rho_{0}, \ldots, \rho_{n}}^{T}$.
In each
Newton step $n$, the linear system of equations
\begin{align}
{\bf J} \cdot {\Delta} \mbox{\boldmath $\rho$} = {\bf
g}(\mbox{\boldmath${\rho}$}^n)
\end{align}
has to be solved, where
the elements of the Jacobian matrix ${\bf J}$ are given by:
\begin{align}
&J_{ij} =\diff{{g_i}}{ \rho_j}  \label{eq:DefJacobian}\\
&=\delta_{ij} \cdot \mu_{\HS}'(\rho_i) +\frac{\Delta z}{2}
\klamm{2 - \delta_{j0} - \delta_{jN}} \cdot  \Phi_{\Pla}(| z_j-
z_i|).\notag
\end{align}

\subsection{Adsorption isotherms \label{eq:ComputationIsotherms}}
{ Solving Eq. (\ref{eq:DiscretizedEquation}) will only give
one density profile $ {\boldsymbol \rho}$ for each chemical
potential $ \mu$. However, in the case of a prewetting transition,
there can be multiple solutions for the same chemical potential.
From these solutions, only one is stable, whereas the other
solutions are meta- or unstable (see also
Sec.~\ref{sec:PlanarIsotherm}). In order to compute the full
bifurcation diagram of the set of density profiles over the chemical
potential, a
pseudo arc-length continuation scheme is
developed similar to the one employed by Salinger and Frink \cite{salinger:7457}.

More specifically, we introduce an arc-length parametrization such
that $\klamm{ \mu(s),{\boldsymbol \rho}(s)}$ with $s \in \mathbb{R}$
is a connected set of solutions of condition
(\ref{eq:DiscretizedEquation}) and where we have included the
chemical potential $\mu$ as an additional variable:
\begin{align}
{\bf g}\klamm{ \mu,{\boldsymbol \rho}} \istobe 0.
\label{eq:Annex_Conti_1DDiscCondition}
\end{align}
The main idea of the continuation scheme is to trace the set of
solutions along the curve parametrized by $s$.

\begin{figure}[htb]
\includegraphics[width=7cm]{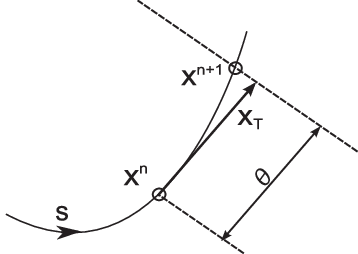}
\caption{Sketch of one iteration step of the continuation scheme.
${\bf x}^n$ and ${\bf x}^{n+1}$ are consecutive points of the
iteration, where, ${\bf x} = \klamm{\mu, {\boldsymbol \rho}}$. ${\bf
x}_T$ is the tangent vector in ${\bf x}^n$. By following the curve
of solutions in the direction of the tangent vector, the pseudo
arc-length continuation scheme is able to trace the curve of
solutions through turning points with respect to the parameter
$\mu$. \label{fig:SktechContinuationMethod}}
\end{figure}

Assume that a point $\klamm{\mu^n,{\boldsymbol \rho}^n}$ at position
$s^n$ on the curve is given, where $n$ is the step of the
continuation scheme being solved for. First, the tangent vector
$\klamm{\frac{d\mu}{ds},\frac{d {\boldsymbol \rho}}{ds}}$ at
position $s^n$ is computed. This is done by differentiating ${\bf
g}(s) \defi {\bf g}\klamm{\mu(s),{\boldsymbol \rho}(s)}$ with
respect to $s$. From (\ref{eq:Annex_Conti_1DDiscCondition}), it is
known that ${\bf g}$ is a constant equal to zero on the curve of
solutions $\klamm{\mu(s),{\boldsymbol \rho}(s)}$. Hence, the
differential $\frac{d{\boldsymbol g}}{ds}$ vanishes:
\begin{align}
\frac{d{\boldsymbol g}}{ds} = \left(
\begin{array}{cc}
\diff{{\boldsymbol g}}{ \mu} & {\bf J}
\end{array}
\right) \cdot \left(
\begin{array}{c}
\frac{d \mu}{ds}\rule[-1.7ex]{0pt}{0pt} \\
\frac{d {\boldsymbol \rho}}{ds}
\end{array}
\right) = 0, \label{eq:Annex_DefiningEquationTangent_general}
\end{align}
where ${\bf J}$ is the Jacobian as defined in (\ref{eq:DefJacobian})
and
\begin{align}
\diff{{g_i}}{ \mu} = -1 + \frac{d \rho_{g}}{d\mu} \Psi_{\Pla}\klamm{ z_N -  z_i}. \label{eq:Annex_Conti_DgDmu}
\end{align}
The second term takes into account that $\rho_{g}$ for the density at $ z >  z_N$ depends on the chemical potential. In our computations, we have approximated $\diff{{g_i}}{ \mu}$ by $-1$.
(\ref{eq:Annex_DefiningEquationTangent_general}) is the defining equation for the tangent vector  $\klamm{\mu_T^n, {\boldsymbol \rho}_T^n} = \klamm{\frac{d\mu}{ds},\frac{d{\boldsymbol \rho}}{ds}}$.

We remark that this homogeneous system of linear equations leaves
one degree of freedom, as we only have $N+1$ equations, but $N+2$
variables, $(\mu_T,{\boldsymbol \rho}_T)$. An additional equation is
then used to maintain the direction of the tangent vector on the
curve of solutions:
\begin{align*}
\left(
\begin{array}{cc}
\mu_T^{n-1} & \klamm{{\boldsymbol \rho}_T^{n-1}}^T
\end{array}
\right) \cdot \left(
\begin{array}{c}
\mu_T^n \\
{\boldsymbol \rho}_T^n
\end{array}
\right) = 1,
\end{align*}
where $\left(
\begin{array}{cc}
\mu_T^{n-1} & \klamm{{\boldsymbol \rho}_T^{n-1}}^T
\end{array}
\right)$ is the tangent vector of the previous iteration.

In a second step, an additional equation for a point at the stepsize
$\theta$ away from $\klamm{ \mu^n,{\boldsymbol \rho}^n}$ and in the
direction of the tangent vector $\left(
\begin{array}{cc}
\mu_T^{n-1} & \klamm{{\boldsymbol \rho}_T^{n-1}}^T
\end{array}
\right)$ is set up. For this purpose we introduce a scalar
product, which takes into account the discretization of the
density profile into $N$ intervals of length $\Delta  z$:
\begin{align}
\langle \klamm{\mu_1 , {\boldsymbol \rho}_1} |  \klamm{\mu_2 , {\boldsymbol \rho}_2}\rangle &\defi \mu_1 \mu_2 + \cdots \\
&\cdots +\frac{\Delta z}{2} \sum_{j=0}^N \klamm{2 - \delta_{j0} -
\delta_{jN}} \cdot \rho_{1j}\rho_{2j}. \notag
\end{align}
The norm with respect to this scalar product is defined as:
\begin{align}
\norm{\klamm{ \mu,{\boldsymbol \rho}}} \defi \langle \klamm{
\mu,{\boldsymbol \rho}}| \klamm{ \mu,{\boldsymbol \rho}}
\rangle^{1/2}.
\end{align}
The curve of solutions $\klamm{ \mu(s),{\boldsymbol \rho}(s)}$ is
now parameterized by the arc-length with respect to the norm given
above, such that,
\begin{align}
\int_{s^n}^{s^n+\theta} \norm{\klamm{\frac{d\mu}{ds},\frac{d
{\boldsymbol \rho}}{ds}}} ds = \theta.
\end{align}
Linearizing the norm around $s^n$ and making use of the
approximate tangent vector $(\mu_T,{\boldsymbol \rho}_T)$ at
$s^n$, one obtains
\begin{align}
\avar{\klamm{ \mu_T^n,{\boldsymbol \rho}_T^n}| \klamm{ \mu(s^n +
\theta) -  \mu(s^n),
 {\boldsymbol \rho}(s^n + \theta) -  {\boldsymbol \rho}(s^n)
} } = \theta, \label{eq:CompIsother_mid1}
\end{align}
where we have made use of the normalized tangent vector such that
\begin{align}
\norm{\klamm{ \mu_T^n,{\boldsymbol \rho}_T^n}} = 1.
\end{align}
Inserting $\klamm{ \mu^{n+1},  {\boldsymbol \rho}^{n+1}}$ for
$\klamm{ \mu(s^n + \theta),  {\boldsymbol \rho}(s^n + \theta)}$ into
(\ref{eq:CompIsother_mid1}) leads to the additional equation for the
next point on the curve of solutions:
\begin{align}
K_n&\klamm{\mu^{n+1},{\boldsymbol \rho}^{n+1}} \defi\label{eq:Annex_Conti_AdditionalEq} \\
&\langle \klamm{ \mu_T^n,{\boldsymbol \rho}_T^n}|
\klamm{\mu^{n+1}-\mu^{n},{\boldsymbol \rho}^{n+1}-{\boldsymbol
\rho}^{n}} \rangle - \theta \quad \istobe \quad 0, \notag
\end{align}
For a geometric interpretation of
Eq.~(\ref{eq:Annex_Conti_AdditionalEq}) see Fig.
\ref{fig:SktechContinuationMethod}.

To obtain $\klamm{\mu^{n+1},{\boldsymbol \rho}^{n+1}}$,
(\ref{eq:Annex_Conti_AdditionalEq}) is solved together with
(\ref{eq:Annex_Conti_1DDiscCondition}). This is done using a Newton
scheme. In each Newton step, the following system of linear
equations is solved:
\begin{align}
\left(
\begin{array}{cc}
\mu_T^n & ({\bar{\boldsymbol\rho}}_T^n)^T\\
\frac{\partial {\bf g}}{\partial\mu} & {\bf J}
\end{array}
\right) \cdot \left(
\begin{array}{c}
\Delta \mu^{m}\\
\Delta {\boldsymbol \rho}^{m}
\end{array}
\right) = \left(
\begin{array}{c}
K_n\klamm{\mu^{n,m},{\boldsymbol \rho}^{n,m}} \\
 {\bf g}(\mu^{n,m},{\boldsymbol \rho}^{n,m})
\end{array}
\right), \label{eq:LinearEqnIsotherm}
\end{align}
where we are considering the $n$-th step of the continuation
scheme and the $m$-th step of the Newton method, such that $\Delta
\mu^{m}\defi \mu^{n,m+1}-\mu^{n,m}$ and $\Delta  {\boldsymbol
\rho}^{m} \defi {\boldsymbol \rho}^{n,m+1}-{\boldsymbol
\rho}^{n,m}$.
Furthermore, we have made use of
\begin{align*}
{\bar{
\rho}}_{T,j}^n \defi \frac{\Delta  z}{2} \klamm{2
- \delta_{j0} - \delta_{jN}}{
\rho}_{T,j}^n.
\end{align*}
Finally, (\ref{eq:LinearEqnIsotherm}) is solved using a conjugate
gradient method, where the Jacobian (\ref{eq:DefJacobian}) of the
system is approximated by introducing a cutoff of $5$ molecular
diameters for the intermolecular potential $\Phi_{\Pla}$.
}

\section{Surface tension and binding potential in the sharp-kink approximation (SKA) \label{sec:AppendixSKA} }
\subsection{Surface tension \label{sex:AppendisSKASurfaceTension}}
According to Gibbsian thermodynamics, the surface tension is the
free energy cost to increase an interface by unit area, i.e. the
excess free energy (excess grand potential for an open system) per
unit area with respect to the corresponding uniform phases. Within
SKA, the liquid-vapour surface tension can be obtained from
(\ref{eq:DefinitionExcessGrandPotential}), with \bb
\rho(\rr)=\left\{\begin{array}{ll}\rho_A,&\rr\in {\mathbb{V}}_A\\
\rho_B&\rr\in {\mathbb{V}}_B\end{array}\right.\,, \ee where
${\mathbb{V}}_A\cap {\mathbb{V}}_B=0$ and ${\mathbb{V}}_A\cup
{\mathbb{V}}_B={\mathbb{R}}^3$. The convenience of the expression
for the excess grand potential as given by
(\ref{eq:DefinitionExcessGrandPotential}) becomes evident now, as
for $\rho_A=\rho_{l}$, $\rho_B=\rho_{g}$ and no external field, only
the second term in (\ref{eq:DefinitionExcessGrandPotential})
matters. One then gets an immediate result for the liquid-gas
surface tension,
 \bb
\gamma_{lg}^{\SKA}=\frac{\Omega^{ex}}{\cal{A}}=-\frac{(\rho_{l}-\rho_{g})^2}{\cal{A}}I({\mathbb{V}}_A,{\mathbb{V}}_B)\,, \label{ap:gamma_ska}
 \ee
where
\begin{align}
I({\mathbb{V}}_A,{\mathbb{V}}_B)\equiv \frac{1}{2}  \int_{{\mathbb{V}}_A}\int_{{\mathbb{V}}_B}\phi(|\rr_1-\rr_2|)\dr_1\dr_2.
\end{align}

For the surface tension of a planar interface we have
$\mathbb{V}_A = \mathbb{V}_{z<0}$ and $\mathbb{V}_B =
\mathbb{V}_{z\geq0}$ such that

$$
\frac{I({\mathbb{V}}_{z<0},{\mathbb{V}}_{z\geq0})}{\cal{A}}= \frac{1}{2}  \int_{-\infty}^0\int_0^{\infty}\Phi_{\Pla}(|z-z'|)dz'dz\,,
$$
with $\Phi_{\Pla}$ defined by (\ref{eq:Definition_Phi1D}). Thus, for
the liquid-gas surface tension we obtain:
\begin{align}
\gamma_{lg}^{\text{SKA}}(\infty)
=& - \frac{\Delta \rho^{2}}{2} \int_{-\infty}^{0} \int_{0}^{\infty}  \Phi_{\Pla}\klamm{|z-z'|} dz' dz \notag\\
=& \frac{3}{4} \pi \Delta \rho^{2} \varepsilon \sigma^{4}. \label{eq:LiqGasSurfaceTensionSKA}
\end{align}

In the case of a spherical symmetry, i.e. a drop of liquid of radius
$R$, $\mathbb{V}_{A} = \{{\bf r}\in \mathbb{R}^{3}: |{\bf r}| < R
\}$ and $\mathbb{V}_{B} = \{{\bf r}\in \mathbb{R}^{3}: |{\bf r}|
\geq R \}$ the surface tension becomes 
 \begin{align}
\gamma_{lg}^{\text{SKA}}(R)=&-\Delta \rho^{2}\frac{I({\mathbb{V}}_{r<R},{\mathbb{V}}_{r\geq R})}{4\pi R^2}\notag\\
=& - \frac{\Delta \rho^{2}}{2} \int_{R}^{\infty} \int_{0}^{R} \klamm{\frac{r}{R}}^{2} \Phi_{\Sph}\klamm{r,r'} dr' dr  \notag\\
=& - \frac{\Delta \rho^{2}}{2} \int_{R}^{\infty}  \klamm{\frac{r}{R}}^{2}   \Psi_{R}\klamm{r} dr  \notag\\
=& \gamma_{lg}(\infty)\klamm{1 - \frac{2}{9} \frac{\ln (R/\sigma)}{(R/\sigma)^{2}} + O\klamm{(\sigma/R)^{2}} }\ \label{ap_gammaR},
 \end{align}
where $\Delta\rho=\rho_l-\rho_g$ and $\Phi_{\Sph}\klamm{r,r'} \equiv \int_{\partial B_{r'}} \phi\klamm{|{\bf r} - {\bf r'}|}d{\bf r'}$ 
can be advantageously expressed in terms of $\Phi_{\Pla}$:

\begin{align}
& \Phi_{\Sph}( r, r') =\notag\\
&= \int_0^{2\pi} \int_0^\pi \phi\klamm{|{\bf  r} - {\bf  r}'|} r'^2 \sin \vartheta ' d \vartheta' d\varphi '\notag\\
&=2\pi r'^{2} \int_0^\pi \phi\klamm{ \sqrt{ r^{2} - 2r r' \cos \vartheta ' + r'^{2} } }
\sin \vartheta ' d \vartheta' \notag\\
&= \pi \frac{r'}{r} \int_{(r-r')^{2}}^{(r+r')^{2}} \phi\klamm{ \sqrt{ t} } dt\notag\\
&= \pi \frac{r'}{r} \klamm{ \int_{(r-r')^{2}}^{\infty} \phi\klamm{ \sqrt{ t} } dt - \int_{(r+r')^{2}}^{\infty} \phi\klamm{ \sqrt{ t} } dt
}\notag\\
&= 2\pi \frac{r'}{r} \left( \int_{0}^{\infty} \phi\klamm{ \sqrt{(r-r')^{2} +u^{2} } } udu \right.
- \notag\\
&\qquad \qquad  \left. \int_{0}^{\infty} \phi\klamm{ \sqrt{(r+r')^{2} +u^{2} } } udu\right)\notag \\
&= \frac{r'}{r}\klamm{ \Phi_{\Pla}\klamm{| r- r'|} -  \Phi_{\Pla}\klamm{| r +  r'|} }, \label{eq:Definition_InteractionSpherical}
\end{align}

and for $r>R$
\begin{align}
&\Psi_{R}( r) \equiv \int_0^{ R} \Phi_{\Sph}( r, r') d r' \label{eq:Spherical_PsiIn}\\
&= \frac{\pi\varepsilon\sigma^4}{3 r} \left\{
\begin{array}{ll}
\frac{\sigma^8}{30}\left[ \frac{ r+9 R}{( r+ R)^9} - \frac{ r-9 R}{( r- R)^9}
\right]+&\\
\qquad +\sigma^2\left[\frac{ r-3 R}{( r- R)^3} - \frac{ r+3 R}{( r+ R)^3}\right]
&   R + \sigma<  r\\
 -{\frac {26}{15}\frac{r}{\sigma} }-\frac{9}{5\sigma^2}\klamm{{
{ R}^{2}- \left(  r-\sigma \right) ^{2}}}+&\\
\quad+{\frac {27}{10}}+\frac{\sigma^8}{30}{\frac { r+9\, R} {\left(
 r+ R \right) ^{9}}}-\sigma^2{\frac { r+3\,
R}{\left(  r+ R
 \right) ^{3}}}
&  r<R +\sigma.
\end{array}\right. \notag
\end{align}
Note that expression (\ref{ap_gammaR}) gives a vanishing Tolman's length.

\subsection{Binding potential}

The binding potential of a system possessing two interfaces is the
surface free energy per unit area of the system minus the
contribution due to the surface tensions of the two interfaces.
It expresses an effective interaction between the interfaces induced
by the attractive forces. If, analogously to the analysis above, we
define three disjoint subspaces ${\mathbb{V}}_W$, ${\mathbb{V}}_A$,
and ${\mathbb{V}}_B$, such that ${\mathbb{V}}_W\cup
{\mathbb{V}}_A\cup {\mathbb{V}}_B={\mathbb{R}}^3$, the density
distribution of the wall-liquid-gas system within SKA is
 \bb
\rho(\rr)=\left\{\begin{array}{ll}0,&\rr\in {\mathbb{V}}_W\\
\rho_{l},&\rr\in {\mathbb{V}}_A\\
\rho_{g}&\rr\in {\mathbb{V}}_B,\end{array}\right.\,,
 \ee
which when substituted into
(\ref{eq:DefinitionExcessGrandPotential}) gives for the excess grand
potential:
\begin{eqnarray}
\Omega_{ex}=&&-\Delta\mu \Delta \rho
{V}_{A}-{\rho_l}^2I({\mathbb{V}}_W,{\mathbb{V}}_{A})-\rho_g^2I({\mathbb{V}}_W,{\mathbb{V}}_B)- \nonumber
\\
&& -(\Delta \rho)^2I({\mathbb{V}}_A,{\mathbb{V}}_B)+\int_{{\mathbb{V}}_A \cup {\mathbb{V}}_B} V(\rr)\rhor \dr . \label{om_ska}
  \end{eqnarray}
We now rearrange the terms in (\ref{om_ska}), such that
  \bb
\frac{\Omega^{ex}(\ell)}{\mathcal{A}}
= -\Delta\mu\Delta \rho \frac{{V}_{A}}{\mathcal{A}}+
\gamma_{wl}^{\SKA}
+ \frac{\mathcal{A}'}{\mathcal{A}}
\gamma_{lg}^{\SKA}+w^{\SKA}(\ell)\,,
  \ee
where $\mathcal{A} = \int_{\partial \mathbb{V}_{W}} dS$ is the surface of the
wall and $\mathcal{A}' = \int_{\partial (\mathbb{V}_{W} \cup \mathbb{V}_{A})} dS$
is the surface of the liquid-gas interface. We obtain
  \begin{align}
 \gamma_{wl}^{\SKA}=&
\frac{1}{\mathcal{A}} \klamm{
-{\rho_l}^2I({\mathbb{V}}_W,{\mathbb{V}}_A \cup{\mathbb{V}}_B)+\rho_l\int_{{\mathbb{V}}_A\cup {\mathbb{V}}_B} V(\rr) \dr},\\
 \gamma_{lg}^{\SKA}=& -\frac{1}{\mathcal{A}'} (\Delta \rho)^2I({\mathbb{V}}_W \cup {\mathbb{V}}_A,{\mathbb{V}}_B),
  \end{align}
and the binding potential $w^{\SKA}$ involving the remaining
contribution
\begin{align}
 w^{\SKA}(\ell)=& \frac{1}{\cal{A}} \klamm{
 2\rho_l \Delta \rho I( \mathbb{V}_W,\mathbb{V}_B)
  -  \Delta \rho \int_{{\mathbb{V}}_B} V(\rr) \dr}.\label{bind}
  \end{align}
Having obtained the expressions of $I(X,Y)$ for systems possessing
translational or spherical symmetry, we can evaluate the binding
potential in the planar case by making use of $\mathbb{V}_{w} =
\mathbb{R}^{2} \times (-\infty,\delta]$, $\mathbb{V}_{A} =
\mathbb{R}^{2} \times (\delta,\ell)$ and $\mathbb{V}_{B} =
\mathbb{R}^{2} \times [\ell,\infty)$:
\begin{align}
w^{\SKA}\klamm{\ell} {\overset{plane}=} &
\Delta \rho \left( \rho_{l} \int_{\ell - \delta}^{\infty} \int_{z}^{\infty} \Phi_{\Pla}\klamm{ z' } dz' dz -\right.\notag\\
& \qquad \qquad  \left. - \int_{\ell}^{\infty} V_{\infty}(z) dz  \right)   \notag \\
=&-\frac{A}{12\pi\ell^2}  \klamm{ 1+ \frac{ 2 + 3 \frac{\delta}{\ell} }{1 - \frac{ \rho_{w}  \varepsilon_{w} \sigma_{w}^{6}  }{ \rho_{l}^{+}  \varepsilon \sigma^{6} }}\frac{\delta}{\ell} +
O\left( \klamm{{\delta}/{\ell}}^{3}\right)}. 
\end{align}
In the spherical case we make use of $\mathbb{V}_{w} = \{  {\bf r}
\in \mathbb{R}^{3}: |{\bf r}| \leq R +\delta  \}$, $\mathbb{V}_{A} =
\{  {\bf r} \in \mathbb{R}^{3}: R + \delta<  |{\bf r}| < R +\ell \}$
and $\mathbb{V}_{B} = \{  {\bf r} \in \mathbb{R}^{3}: |{\bf r}| \geq
R + \ell \}$ to obtain
\begin{align}
w^{\SKA}\klamm{\ell;R} &{\overset{sphere}=}
=w^{\SKA}\klamm{\ell;\infty}\left(1+\frac{\ell}{R}\right)
\label{eq:BindingSphereSKAAppendix}
\end{align}
where we have neglected terms
$O\klamm{\klamm{\delta/\ell}^{3},\delta/R,\frac{  \ln
(\ell/R)}{(R/\ell)^{2}}}$.

\section{Surface tension, binding potential, and the Tolman length in the soft-interface approximation (SIA)}
\subsection{Surface tension \label{sex:AppendixSIASurfaceTension}}
The surface tension of a planar liquid-gas interface in the
soft-interface approximation
\begin{align}
\rho_{lg, \infty}(z) = \left\{\begin{array}{ll}
\rho_{l} &  z \leq  -\chi/2\\
\rho_{lg}(z)  & |z|< \chi/2\\
\rho_{g} &  z \geq  \chi/2
\end{array}\right.,
\label{sia_planar}
\end{align}
is obtained by substituting (\ref{sia_planar}) into (\ref{eq:DefinitionExcessGrandPotential}) with $V(\rr)=0$
\begin{align}
 \gamma_{lg}^{\SIA}(\infty)&= \frac{\Omega_{ex} [\rho_{lg,\infty}] }{\mathcal{A}}\label{eq:SIASurfaceTensionPlanar}\\
 &= - \int_{-\chi/2}^{\chi/2} \klamm{ p( \rho_{lg,\infty}( z))-  p(\rho_{ref}( z))}   d z+ \notag\\
&+ \frac{1}{2}\int_{-\infty}^\infty \int_{-\infty}^\infty \rho_{lg,\infty}( z) \klamm{\rho_{lg,\infty}( z') -\rho_{lg,\infty}( z)}\times \notag\\
& \qquad \times \Phi_{\Pla}( z, z') d z' d z\notag,
\end{align}
where $\rho_{ref}(z)$ denotes the density of a given bulk phase, \newstuff{i.e.
$\rho_{ref}(z)=\rho_l\Theta(-z)+\rho_g\Theta(z)$ such that at saturation
$p\klamm{\rho_{ref}(z)} = p_{ref} = \text{const}$}. We note that in
the above approximation the contribution due to the excess local
pressure is generally non-zero (in contrast to the SKA).

In the spherical case, the density profile is
\begin{align}
\rho_{lg, R}( r) =  \left\{\begin{array}{ll}
\rho_{l} &  r \leq  R-\chi/2\\
\rho_{lg}( r- R)  & | r- R|< \chi/2\\
\rho_{g} &  r \geq R+\chi/2
\end{array}\right.,
\label{sia_sphere}
\end{align}
and the surface tension of a liquid drop of radius $R$ is
 \begin{align}
 \gamma_{lg}^{\SIA}&(R)= \frac{\Omega_{ex} [\rho_{lg,R}] }{4\pi R^{2}}\label{eq:SIASurfaceTensionCuved}\\
 &= - \int_{R-\chi/2}^{R+\chi/2} \klamm{ p( \rho_{lg,R}( r))-  p_{ref}}  \klamm{\frac{r}{R}}^2 d r+ \notag\\
&+ \frac{1}{2} \int_0^\infty \int_0^\infty\rho_{lg,R}( r) \klamm{\rho_{lg,R}( r') -\rho_{lg,R}( r)}\times \notag\\
& \qquad \times \Phi_{\Sph}( r, r')  \klamm{\frac{r}{R}}^2 d r' d r.\notag
\end{align}

\subsection{Tolman length \label{sec:AppendixSIATolman}}
Here we calculate the Tolman length as given by SIA by a direct
comparison of (\ref{eq:SIASurfaceTensionPlanar}) and
(\ref{eq:SIASurfaceTensionCuved}). We first compare the second terms
of (\ref{eq:SIASurfaceTensionPlanar}) and
(\ref{eq:SIASurfaceTensionCuved}). For this purpose we define
\begin{align}
h_{R}(r,r')   &\equiv\rho_{lg, R}( r) \klamm{\rho_{lg, R}( r') -\rho_{lg, R}( r)}\notag \\
\text{and} \qquad
h(r,r')   &\equiv\rho_{lg,\infty}( r) \klamm{\rho_{lg,\infty}( r') -\rho_{lg,\infty}( r)}
\end{align}
and taking use of (\ref{eq:Definition_InteractionSpherical}) we can
express the double integral in (\ref{eq:SIASurfaceTensionCuved}) as
\begin{align}
&
\frac{\sigma^{2}}{\varepsilon}
\int_0^\infty\int_0^\infty h_{R}(r,r') \Phi_{\Sph}( r, r')  \klamm{\frac{r}{R}}^2 d r' d r \notag\\
& = \frac{\sigma^{2}}{\varepsilon} \int_{-R}^{\infty}\int_{-R}^{\infty} h(r,r')  \left( \Phi_{\Pla}( |r-r'| ) - \right. \notag\\
&\qquad
\left. -
 \Phi_{\Pla}( |2R+ r-r'| ) \right) 
\klamm{1+\frac{r'}{R}}  \klamm{1+\frac{r}{R}} dr' dr \notag\\
& = \frac{\sigma^{2}}{\varepsilon} \int_{-R}^{\infty}\int_{-R}^{\infty} h(r,r') \Phi_{\Pla}( |r-r'| )  \times\notag\\
&\qquad \times \klamm{1+\frac{r'}{R}}  \klamm{1+\frac{r}{R}} dr' dr + O\klamm{(\sigma/R)^{2}} \notag\\
& =  \frac{\sigma^{2}}{\varepsilon} \int_{-R}^{\infty} \int_{-R}^{\infty}h(r,r') \Phi_{\Pla}( |r-r'| ) \times \notag\\
&\qquad  \qquad \times \klamm{1+\frac{r+r'}{R}}  dr' dr  +
O\klamm{  \frac{ {\ln (R/\sigma)} }{ (R/\sigma)^{2} }  }
\notag \\
& =  \frac{\sigma^{2}}{\varepsilon} \int_{-\infty}^{\infty}  \int_{-\infty}^{\infty} h(r,r') \Phi_{\Pla}( |r-r'| ) \times \notag\\
& \qquad \qquad  \times \klamm{1+\frac{r+r'}{R}}  dr' dr +
O\klamm{  \frac{ {\ln (R/\sigma)} }{ (R/\sigma)^{2} }  }
\label{eq:TolmanLengthFirstStep}.
\end{align}
Comparison with the double integral in
(\ref{eq:SIASurfaceTensionPlanar}) then yields
\begin{align}
& \frac{\sigma^{2}}{\varepsilon R} \iint_{-\infty}^{\infty} h(r,r') \Phi_{\Pla}( |r-r'| ) \klamm{r+r'}  dr' dr  +\notag\\
& \qquad+ O\klamm{  \frac{ {\ln (R/\sigma)} }{ (R/\sigma)^{2} }  }   \notag\\
&= \frac{\sigma^{2}}{\varepsilon R} \int_{-\infty}^{\infty} \int_{-\infty}^{\infty}  r \klamm{ h(r,r') + h(r',r)} \Phi_{\Pla}( |r-r'| ) dr' dr+ \notag\\
& \qquad \qquad +O\klamm{  \frac{ {\ln (R/\sigma)} }{ (R/\sigma)^{2} }  }  \notag\\
&= - \frac{\sigma^{2}}{\varepsilon R} \int_{-\infty}^{\infty} \int_{-\infty}^{\infty} r \klamm{\rho_{lg, \infty}( r') -\rho_{lg, \infty}( r)}^{2} \times \notag\\
& \qquad \qquad  \times  \Phi_{\Pla}( |r-r'| ) dr' dr + O\klamm{  \frac{ {\ln (R/\sigma)} }{ (R/\sigma)^{2} }  } .  \notag
\end{align}
\newstuff{In the following, we focus on the asymmetry of the model due to the contribution of the pressure but for simplicity
we assume that the density profile is symmetric. In this case, the
integrand in the above expression is anti-symmetric with respect to
the reflection transformation $r\rightarrow-r$ and
$r'\rightarrow-r'$ and the term $O(\sigma/R)$ vanishes.}

For the difference of the first terms of (\ref{eq:SIASurfaceTensionCuved}) and (\ref{eq:SIASurfaceTensionPlanar}) we obtain
\begin{align}
&-  \frac{\sigma^{2}}{\varepsilon} \int_0^\infty \klamm{ p( \rho_{lg,R}( r))-  p_{ref}}  \klamm{\frac{r}{R}}^2 d r \notag\\
&+ \frac{\sigma^{2}}{\varepsilon} \int_{-\infty}^\infty \klamm{ p( \rho_{lg,\infty}( z))-  p_{ref}} dz\notag\\
&= - \frac{2 \sigma^{2}}{ \varepsilon R} \int_{-\chi/2}^{\chi/2} \klamm{ p( \rho_{lg,\infty}( z))-  p_{ref}}  z d z+ \notag\\
&\qquad \qquad \qquad+ O\klamm{(\sigma/R)^{2}}\,,
\end{align}
yielding a Tolman length
 \bb
 \delta_{\infty} = \frac{1}{\gamma_{lg}^{\SIA}(\infty)} \int_{-\chi/2}^{\chi/2} \klamm{ p(\rho_{lg}(z)) - p_{ref}} z dz.
 \ee
 \newstuff{Note that in line with \cite{Fisher:1984kx}, the Tolman length does not depend on the choice of the dividing surface.}

\subsection{Binding potential \label{sec:AppendixSIABindingPot}}
The extension of the expression for the binding potential, Eq.~(\ref{eq:BindingSphereSKAAppendix}), as given by SKA is rather straightforward. We consider the density distribution as follows
 \bb
\rho(\rr)=\left\{\begin{array}{ll}0 &\rr\in {\mathbb{V}}_W\\
\rho_l&\rr\in {\mathbb{V}}_A\\
\rho_{lg}(\rr),&\rr\in {\mathbb{V}}_{AB}\\
\rho_g&\rr\in {\mathbb{V}}_B,\end{array}\right.\,,
 \ee
for ${\mathbb{V}}_W$ a sphere of radius $R+\delta$,
${\mathbb{V}}_W\cup{\mathbb{V}}_A$  a sphere of radius
$R+\ell-\chi/2$,
${\mathbb{V}}_W\cup{\mathbb{V}}_A\cup{\mathbb{V}}_{AB}$ a sphere of
radius $R+\ell+\chi/2$ and ${\mathbb{V}}_W\cup {\mathbb{V}}_A\cup
{\mathbb{V}}_B\cup {\mathbb{V}}_{AB}={\mathbb{R}}^3$. Such a model
is relevant for the study of wetting on a spherical ($R$ finite) and
on a planar ($R\rightarrow\infty$) wall. It should be noted that in
contrast to SKA, this density distribution is not piecewise
constant, due to the position dependent part of $\rhor$ in the
region ${\mathbb{V}}_{AB}$. Furthermore, we define the following
operators
\begin{align*}
[XY] &\equiv -\frac{1}{2} \int_{X}\int_{Y} \klamm{\rho\klamm{\bf r} -\rho\klamm{ \bf r'}}^{2} \phi\klamm{ | {\bf r} - {\bf r'} | } d{\bf r}' d {\bf r}\\
[XY]_{wl} &\equiv -\frac{1}{2} \int_{X}\int_{Y} \klamm{\rho_{wl}\klamm{\bf r} -\rho_{wl}\klamm{ \bf r'}}^{2} \phi\klamm{ | {\bf r} - {\bf r'} | } d{\bf r}' d {\bf r}\\
[XY]_{lg} &\equiv -\frac{1}{2} \int_{X}\int_{Y} \klamm{\rho_{lg}\klamm{\bf r} -\rho_{lg}\klamm{ \bf r'}}^{2} \phi\klamm{ | {\bf r} - {\bf r'} | } d{\bf r}' d {\bf r}\,,\\
\end{align*}
with $\rho_{wl}\klamm{\bf r}\equiv\rho_l\chi_{{\mathbb{R}}^3\setminus{\mathbb{V}}_W}(\rr)$ and $\rho_{lg}\klamm{\bf r} \equiv \rho_l\chi_{{\mathbb{V}}_W\cup{\mathbb{V_A}}}(\rr) +\rho_{lg}(\rr)\chi_{{\mathbb{V}}_{AB}}(\rr)
+\rho_g\chi_{\mathbb{V}_B}(\rr) $, where $\chi_X(\rr)$ is the characteristic function of a subset X.
Using this convention, the wall-liquid and liquid-gas surface
tensions can be respectively expressed as
\begin{align*}
\gamma_{wl} =& \frac{1}{\cal{A}}\left([\mathbb{V}_W\mathbb{V}_{A}] + [ \mathbb{V}_W \klamm{ \mathbb{V}_{AB} \cup \mathbb{V}_{B}}]_{wl} + \right. \notag\\
& \left. \qquad \qquad
+ \int\rho_{wl}({\bf r}) V(\rr) d\rr  \right)\\
\gamma_{lg} =&  \frac{1}{\cal{A}}\left([\mathbb{V}_{AB}\mathbb{V}_{B}] + \frac{1}{2} [ \mathbb{V}_{AB}\mathbb{V}_{AB}] + [ \mathbb{V}_{A} \klamm{ \mathbb{V}_{AB} \cup \mathbb{V}_{B}}]+\notag\right.\\
& \qquad  +[ \mathbb{V}_W \klamm{ \mathbb{V}_{AB} \cup \mathbb{V}_{B}}]_{lg}  -  \notag\\
&\qquad \left.- \int_{\mathbb{V}_{AB} }   \klamm{   p(\rho_{lg}\klamm{\bf r}   )  -  p_{ref} }    d{\bf r} \right), \label{sia_gamma}
\end{align*}
where $\mathcal{A}=4\pi R^{2}$.
When this is subtracted from the
surface grand potential (\ref{eq:DefinitionExcessGrandPotential}),
which can be written as
\begin{align}
&\frac{\Omega_{ex} }{\mathcal{A}} = \frac{1}{\mathcal{A}}\left( [\mathbb{V}_{W}\mathbb{V}_{A}]+
[\mathbb{V}_{W}(\mathbb{V}_{AB}\cup \mathbb{V}_{B})]+\right.\notag\\
&\quad +[\mathbb{V}_{A}(\mathbb{V}_{AB}\cup \mathbb{V}_{B})]
+ \frac{1}{2} [\mathbb{V}_{AB}\mathbb{V}_{AB} ] +
[\mathbb{V}_{AB}\mathbb{V}_{B}]
-\notag\\
& \quad- \int_{\mathbb{V}_{AB} }   \klamm{   p(\rho_{lg}\klamm{\bf r}   )  -  p(\rho_{ref}\klamm{\bf r}   )  }    d{\bf r}+\notag\\
&\left. \quad+ \int \rho ({\bf r}) V(\rr) d\rr  \right),
\end{align}
one obtains for the binding potential:
\begin{align}
&w^{\SIA} =\frac{1}{\cal{A}}\left([ \mathbb{V}_W \klamm{ \mathbb{V}_{AB} \cup \mathbb{V}_{B}}]  - [\mathbb{V}_W  \klamm{ \mathbb{V}_{AB} \cup \mathbb{V}_{B}}]_{wl}  -  \right.   \notag\\
 &\left.- [ \mathbb{V}_W \klamm{ \mathbb{V}_{AB}\cup \mathbb{V}_{B}}]_{lg}
 + \int V\klamm{\bf r} \klamm{   \rho\klamm{\bf r}  - \rho_{wl}\klamm{\bf r} d{\bf r}  }
 \right). \notag
\end{align}
In spherical coordinates, the binding potential reads:
\begin{align}
w^{\SIA} & =\int_{0}^{R+\delta} \int_{R+\ell-\chi/2}^{\infty} \left(\frac{r}{R}\right)^2
\rho_{l}\klamm{\rho_{l}-\rho(r')}\times \notag\\
&\qquad\qquad\times  \Phi_{\Sph}\klamm{ r,r'} d{r'} d{r} + \notag\\
&\qquad + \int_{R+\ell -\chi/2}^{\infty} \klamm{ \rho(r)-\rho_l } V_{R}(r) \left(\frac{r}{R}\right)^2 dr.
\end{align}

\bibliography{EC11023_BIB}

\end{document}